\documentclass[%
 reprint,
superscriptaddress,
showpacs,preprintnumbers,
 amsmath,amssymb,
 aps,
]{revtex4-1}

\usepackage{latexsym}
\usepackage{amsmath}
\usepackage{float}

\usepackage{soul}
\usepackage{ulem}
\usepackage{graphicx}
\usepackage{dcolumn}
\usepackage{bm}
\usepackage{color}
\newcommand\redout{\bgroup\markoverwith{\textcolor{red}{\rule[.5ex]{2pt}{0.4pt}}}\ULon}

\usepackage[colorlinks=true,citecolor=blue]{hyperref}
\hypersetup{colorlinks=true,citecolor=blue,linkcolor=red,urlcolor=blue}

\begin{document}
\date{\today}

\title{Strain-induced modulation of exchange interaction in monolayer zigzag nanoribbons of B$_2$S }
\author{Moslem Zare }
\email{mzare@yu.ac.ir}
\affiliation{Department of Physics, Yasouj University, Yasouj, Iran 75914-353, Iran}

\date{\today}

\begin{abstract}
In this work, the strain modulation of electronic structure and magnetic interaction in monolayer nanoribbons of B$_2$S, a recently realized monolayer system, is investigated.
In the first part of our study, we focus on how the electronic structure of monolayer nanoribbons of B$_2$S is modified under uniaxial strains, then  employing a tight-binding framework together with the conventional theory of elasticity, we discuss how strain-induced local deformations can be used as a means to affect Ruderman-Kittel-Kasuya-Yosida (RKKY) interaction in zigzag nanoribbons of B$_2$S.
We show that breaking inversion symmetry in zigzag B$_2$S nanoribbons (ZBSNR), e.g., by introducing staggered sublattice potentials, play a key role in the modulation of their electronic properties.
More interestingly, for the ZBSNRs belong to the group $M=4p$, with $M$ the width of the ZBSNR and $p$ an integer number, one can see that a band gap, in which a pair of near-midgap bands completely detached from the bulk bands is always observed.
As a key feature, the position of the midgap bands in the energy diagram of ZBSNRs can be shifted by applying the in-plane strains $\varepsilon_x$ and $\varepsilon_y$. Moreover, the near-midgap bandwidth monotonically decreases with increasing strength of the strain and increases with the width of the ZBSNR. The energy gap of the ZBSNRs decreased with increasing the applied strain and ribbon width.
The spatial and strain dependency of the exchange interaction in various configurations of the magnetic impurities are also evaluated.
It is shown that that magnetic interactions between adsorbed magnetic impurities in B$_2$S nanoribbons can be manipulated by careful strain engineering of such systems. Our results suggest that these tunable electronic and magnetic properties of ZBSNRs mean they may find applications in spintronics and pseudospin electronics based on monolayer B$_2$S nanoribbons.

\end{abstract}

\maketitle

\section{Introduction}
The Ruderman-Kittel-Kasuya-Yosida (RKKY) interaction \cite{Ruderman, Kasuya, Yosida}, is an indirect exchange coupling between magnetic impurity dopants mediated by a background of conduction electrons of the host material. Of fundamental interest in the field of spintronics, the RKKY interaction is the most important mechanism of the coupling between localized spins in metals and semiconductors. Depending on the spatial separation of the magnetic impurities, the magnetic coupling could be ferromagnetic \cite{Vozmediano, Brey, Priour, Matsukura, Ko, Ohno-science} or antiferromagnetic \cite{Minamitani, Loss15} and is oscillatory due to the sharpness of the Fermi surface. Another interesting phases such as spiral \cite{moslem-si, Mahroo-RKKY} and spin-glass \cite{pesin, Eggenkamp, Liu87}, are also attainable in the magnetically doped systems. Besides these practical magnetic phases, the RKKY interaction can provide information about the intrinsic properties of the material since this coupling is proportional to the spin susceptibility of the host system.
A key feature of the RKKY exchange is the long-range oscillations with the Fermi wavevector originates from the Friedel oscillations, that falls off by $R^{-D}$, where $D$ is the dimension of the system \cite{fariborz-blg, fariborz-mos2}. In undoped single layer graphene, the RKKY interaction has contributions that decay with $R^{-3}$, a reflection of the vanishing density of states at the Fermi energy, while it falls off as $R^{-2}$ in doped case.
It was shown that the RKKY interaction consists of three Heisenberg, Ising, and Dzyaloshinskii–Moriya (DM) terms on the surface of zigzag silicene nanoribbons as well as the three dimensional topological insulators~\cite{pesin,moslem-si,JJZhu}, and the competition between them causes rich spin textures.

An additional term, namely spin-frustrated has discovered in a three-dimensional Weyl semimetal (WSM) that along with the Dzyaloshinskii-Moriya term lies in the plane perpendicular to the line connecting two Weyl points~\cite{M.V.Hosseini}. A decay rate of $R^{-3}$ for doped WSM is found, while a much faster decay rate of $R^{-5}$ is found for half-filling band at Weyl points.

In a spin polarized system~\cite{fariborz-sp} and a material with multi-band structure~\cite{fariborz-mos2}, these oscillations become more complicated than a monotonic oscillation with $\sin(2k_{\rm F}R)$ behavior, where $k_{\rm F}$ is the wave vector of the electrons (holes) at the Fermi level and $R$ is the distance of two magnetic impurities. In addition, it is important to note that the magnitude of the RKKY interaction can be severely affected by the density of states (DOS) at the Fermi level~\cite{fariborz-blg, Mahroo-RKKY}. Owing to the bipartite nature of the honeycomb sublattice, the RKKY coupling in graphene is highly sensitive to the direction of the distance vector between impurities~\cite{sherafati-g, fariborz-blg}. In materials with spin-orbit interaction of Rashba type~\cite{Mahroo-single}, the exchange interaction depends on the direction of the magnetic moments and as the result the RKKY interaction becomes anisotropic~\cite{Mahroo-RKKY}.
Recently, in a detailed study it has shown that the topological phase transition in the zigzag silicene nanoribbons can be probed by using the RKKY interaction\cite{moslem-si}. In another work, it has concluded that the RKKY interaction in the bulk phosphorene monolayer is highly anisotropic and the magnetic ground-state of two magnetic adatoms can be tuned by changing the spatial configuration of impurities as well as the chemical potential varying \cite{Moslem18}.

In the last decades, dilute semiconductors have emerged as a research hotspot due to their functionalities for application in spintronic devices and magnetic recording media~\cite{fabian, Babar,W.Han}.
In this regard, inducing magnetism in otherwise nonmagnetic two dimensional (2D) materials may be lead to next generation of spintronic devices based on the spin degrees of freedom~\cite{pesin,moslem-si,JJZhu}. Motivated by the interaction of two dimensional lattices with magnetic objects, we have recently addressed the problem of indirect exchange coupling between localized magnetic moments mediated by the conduction electrons of 2D materials~\cite{moslem-si,Moslem18,MoslemBP}. Of particular interest is the potential for zigzag nanoribbon-based spintronic devices to be realized, and thus much attention has been focused on determining the magnetic properties of 2D honeycomb nanoribbons. The RKKY interaction in nanoribbon of two dimensional lattices has attracted strong attention in condensed matter physics~\cite{Klinovaja13,moslem-si,Moslem18,Duan17}.

On the other hand, strain engineering, a key strategy for manipulating the magnetic coupling in 2D nanostructures~\cite{Duan17}, has a perfect platform for its implementation in the atomically thin materials. Motivated by the search for materials for spintronics, a huge number of works have been performed to examine the effectiveness of mechanical strain in modulating the magnetic properties of 2D layered materials~\cite{Pereira09,PhysRevB.76.064120, pereira_tight-binding_2009,ourpaper,Peng20123434, Guinea:gapsgraphene,sharma_effect_2013,Duan17}

Very recently, a new 2D anisotropic Dirac cone material, B$_2$S monolayer, appears in the research field again, by using global structure search method and first principles calculation combined with tight-binding model~\cite{Yu.Zhao,P.Li}. B$_2$S monolayer, showing a Fermi-velocity of 106 $m/s$ in the same order of magnitude as that of graphene, was found to be mechanically, thermally and dynamically stable. It is the first pristine honeycomb lattice with a tilted anisotropic Dirac cone structure, stabilized by sulfur atoms in free standing condition. Since, boron atom has one electron less than carbon all the reported 2D boron-based Dirac cone materials, have much more complicated geometries in comparison with the pristine honeycomb structure of graphene.
Both theoretical and experimental studies have shown that the pristine honeycomb geometry of 2D boron sheet is unstable in its freestanding form which can be stabilized by adding two electrons to each boron hexagonal ring via doping metal atoms~\cite{L.Z.Zhang,H.Zhang}.

In this paper, based on the Green's function technique, within the tight-binding model we investigate the RKKY interaction between two magnetic impurities placed on the same/different sublattices of B$_2$S nanoribbon. In the first part of our study, we focus on how the electronic structure of monolayer nanoribbons of B$_2$S is modified under uniaxial strains as well as by introducing staggered sublattice potentials, then employing a tight-binding approach together with the conventional theory of elasticity, we discuss how strain-induced local deformations can be used as a means to affect magnetic exchange interaction in zigzag nanoribbons of B$_2$S.
Morevere, we show that for the ZBSNRs belong to group $M=4p$, with $M$ the width of the ZBSNR ($M$ atoms wide) and $p$ an integer number, a band gap, in which a pair of near-midgap bands completely detached from the bulk bands, is always observed.

This paper is organized as follows. In Sec.~\ref{sec:theory}, we introduce the system under consideration, i.e., a monolayer zigzag B$_2$S nanoribbon
under the influence of strain and staggered potential applied to it. A tight-binding model Hamiltonian for monolayer B$_2$S is presented and then the band spectrum of ZBSNR under a staggered potential is calculated. Then we introduce the theoretical framework which will be used in calculating the RKKY interaction from the real space Green’s function. After that, we discuss our numerical results for the proposed magnetic doped ZBSNR in the presence of a in-plane strain. Finally, our conclusions are summarized in Sec.~\ref{sec:summary}.

\begin{figure}
\includegraphics[width=8.5cm]{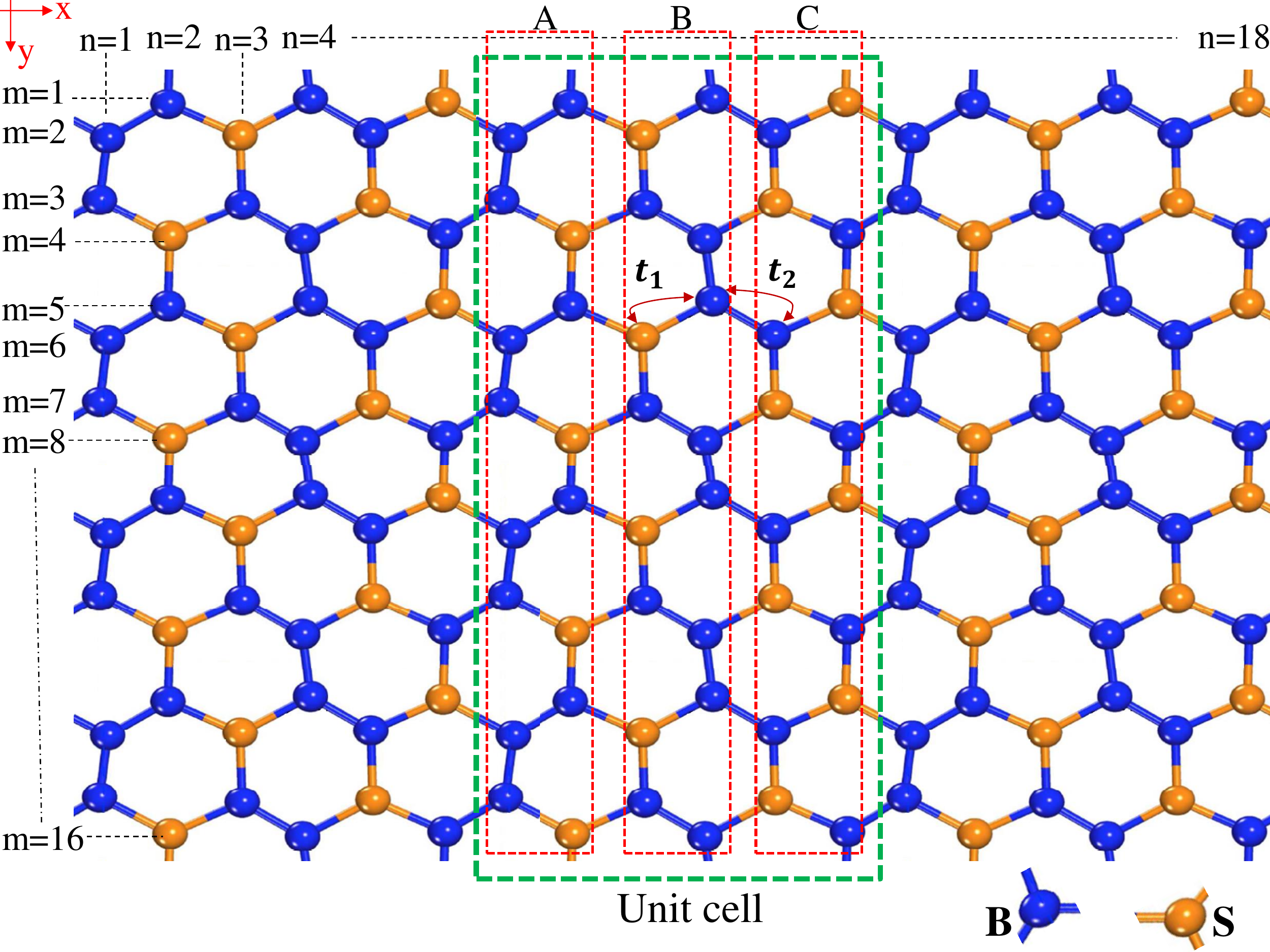}
\caption{(Color online) Schematic illustration of a zigzag B$_2$S nanoribbon of width $M$ ($M$ atoms wide {\it i.e.}, number of zigzag chains), and length $L=N$, where $N$ is the number of vertical armchair chains of type A, B or C, as depicted. The zigzag edge is at the $x$ direction. The green-dashed rectangle denotes the unit cell of the system in which the red-dashed rectangles denote three kind of the vertical in each unit cell, namely A, B, C. B corresponds to the Boron atom, and S to the Sulfur atoms. The nearest-neighbor hopping parameters, used in the tight-binding Eq.~\ref{eqn:TBhamiltonian}, are denoted by $t_1$ and $t_2$. For simplicity, each atom is labeled with a set $(n, m)$, where $n$, $m$ represent the $x$ and $y$ coordinates of the lattice sites. }
\label{fig2}
\end{figure}

\section{Theory and model}\label{sec:theory}

As previously mentioned, the most stable atomic configuration of the B$_2$S monolayer, as a strained graphene, has a honeycomb structure in which each hexagonal ring is distorted with the bond angles ranging from $114$ \AA~ to $123$ \AA, because B and S atoms have different covalent radii and
electronegativities~\cite{Yu.Zhao,,P.Li}.
The geometry structure of a monolayer zigzag nanoribbon of B$_2$S, laid in the $xy$ plane, is depicted in Fig.\ref{fig2}. As shown in this figure, each hexagon consists of four B atoms and two S atoms, with an orthogonal primitive cell with a space group of $PBAM$ and a point group $D_2h$. In this structure, there are two kinds of bonds with $1.62$ \AA  and $1.82$ \AA  bond lengths for B-B and B-S connections, respectively~\cite{Yu.Zhao}.

Moreover, for better strain engineering in 2D layered materials, a deeper understanding of how the geometry and
the electronic structure change by imposing lattice strains on the samples. To verify the role of strain in the magnetic exchange interaction, the understanding of changes in the band structure and bandgap transformation is crucial.
For B$_2$S monolayer, it is shown that the bands near the Fermi level are originated from the $p_z$ orbitals~\cite{Yu.Zhao}. A ‎nearest-neighbor effective tight-binding Hamiltonian, in the basis of $p_z$ orbitals and in the second quantized representation has recently been modeled ~\cite{Yu.Zhao} and is given by‎

‎\begin{equation}‎
‎H_{B_2 S} =\sum_{i}U_{i} c_{i}^\dagger c_{i}‎+ ‎\sum_{\langle i,j\rangle} t_{i,j}‎
‎c_{i}^{\dagger}c_{j}‎,
‎\label{eqn:TBhamiltonian}‎
‎\end{equation}‎
where $c_{i}^{\dagger}$ ($c_{j}$) represents the creation (annihilation) operator of electrons at site $i (j)$, ‎$U_{i}$ is the onsite energy of‎ ‎the $i$-th atom and $t_{i,j}$ is the nearest-neighbor hopping amplitude‎ ‎between $i$-th atom and $j$-th atom‎. $\langle ~ \rangle$ denotes the nearest neighbors. The suggested values of these hopping integrals are specified as $t_{1}=0.8$ eV and $t_{2}=1.7$ eV ~\cite{Yu.Zhao}.
However, it has been found that the onsite energies for S and B atom are 6.4 eV and 5.4 eV, respectively which means that the sulfur atom is more electronegative {\it i.e.}, it is better than boron at attracting electrons.

Having an accurate tight-binding model, as presented in the previous equation, we can numerically calculate the momentum space dispersion of a monolayer zigzag nanoribbon of B$_2$S directly. To do so, we assume the periodic boundary condition along the ribbon in the $x-$direction.

Owing to the translational invariant along the ribbon direction ($x$), the momentum in the $x-$direction is a good quantum number thus the $k$-dependent band structure of ZBSNR, from the tight-binding model, is obtained from $\sum_{{\bm k}}\psi_{{\bm k}}^{\dag} H_k \psi_{{\bm k}}$. In order to obtain the $k$-space Hamiltonian $H_k$, one can, e.g., perform the Fourier transformation along the $x-$direction, to the real space Hamiltonian, written as
\begin{equation}\label{eq:H-k1}
 H_k=H_{00}+H_{01}e^{-i k_x
a}+H_{01}^\dagger e^{i k_x a}
\end{equation}
in which $a$ is the unit-cell length along the x-axis. Moreover, $H_{00}$ and $H_{01}$ describe coupling within the principal unit cell (intra-unit cell) and between the adjacent principal unit cells (inter-unit cell), respectively based on the real space tight-binding model given by Eq.~\ref{eqn:TBhamiltonian}‎.
The corresponding intercellular part of Hamiltonian $H_{00}$, can be written as

‎\begin{equation}‎
H_{00}=\left(
‎\begin{array}{ccccc}‎
H_{AA} & H_{AB}  & H_{AC}  \\‎
H_{BA} & H_{BB}  & H_{BC} \\‎
H_{CA} & H_{CB}  & H_{CC} \\‎
‎\end{array}
‎\right)‎ ,
‎\end{equation}

and the intracellular coupling Hamiltonian between two adjacent unit cells, $H_{01}$, is written in the form
‎\begin{equation}‎
H_{01}=\left(
‎\begin{array}{ccccc}‎
H_{A_1A_2} & H_{A_1B_2}  & H_{A_1C_2}  \\‎
H_{B_1A_2} & H_{B_1B_2}  & H_{B_1C_2} \\‎
H_{C_1A_2} & H_{C_1B_2}  & H_{C_1C_2} \\‎
‎\end{array}
‎\right)‎ ,
‎\end{equation}

To explain the band structure of the nanoribbons, the eigenvalues of the Hamiltonian~\ref{eq:H-k1} must be solved.
Furthermore, corresponding wave function for a given energy and wave vector can be used in order to evaluate the site-resolved local density of states (LDOS) in the ribbons as $\rho({\bf r},E_{nk})=\sum_{m k'}{|\psi_{m k'}({\bf r})|^2 \delta(E_{n k}-E_{m k'})}$, where $n,m$ are band indexes and ${\bf r}$ is the sublattice position.

\subsection{Inclusion of strain}
In this subsection, the effect of strain on the band structure and magnetic exchange interaction is analyzed and discussed.
We first consider a zigzag B$_2$S nanoribbon lattice in the $xy$ plane, in the presence of uniaxial strains $\epsilon_x$ and $\epsilon_y$.
As is known, the strain modulates the band structure by modifying the hopping parameters and hence a significant influence on the magnetic exchange interaction is expected.
As it has been shown, the strength of the hopping parameter ($t$) between $s$ and $p$ orbitals depends very strongly on the bond length ($r$) and can be written as $t\propto\frac{1}{r^{2}}$~\cite{HarrisonWA1999,TangH,J.W.Jiang}. It is worthwhile to note that since it has been assumed that the principal directions of the two neighboring Wannier orbitals keep their orientation along the bond vector of the two neighbor sites, the angular dependence of the mechanical strain is negligible. Moreover, the applied mechanical strain can affect the electronic states through modifying the hopping parameters in the tight-binding model.

Let the $x$-axis be in the direction of the zigzag edge of B$_2$S nanoribbon and the y-axis in that of the armchair edge, as seen in Fig.\ref{fig2}. Within the context of continuum mechanics and in the linear deformation regime, application of a uniaxial strain will cause the following change of the bond length $r$, in terms of strain components $\epsilon_{x}$, and $\epsilon_{y}$

\begin{eqnarray}
\left(\begin{array}{c}
x'\\
y'
\end{array}\right) & = & \left(\begin{array}{ccc}
1+\epsilon_{x} & \gamma \\
\gamma & 1+\epsilon_{y}
\end{array}\right)\left(\begin{array}{c}
x\\
y
\end{array}\right),
\end{eqnarray}
where ${\bf r}=x {\bf i}+ y {\bf j}$ and ${\bf r'}=x' {\bf i}+ y' {\bf j}$ denote the positions of an atom before and after deformation, respectively.

In the linear deformation regime, an expansion of the norm of $r$ to first order in strains $\epsilon_x$ and $\epsilon_y$ can be
expressed as

\begin{equation}
  r'\simeq(1+\alpha_x\epsilon_x+\alpha_y\epsilon_y) r,
 \label{rstrain}
\end{equation}
where $\alpha_x={({x}/{r})}^2$ and $\alpha_y={({y}/{r})}^2$ are the strain-related geometrical coefficients in zigzag B$_2$S nanoribbon.
By invoking the relationship between the hopping parameter and the bond length, we get the following geometrical strain effect on the hopping parameter,
\begin{eqnarray}
t & = & t_{0}\left(1-\frac{2}{r}\alpha_{x}\epsilon_{x}-\frac{2}{r}\alpha_{y}\epsilon_{y}\right).
\label{eq_t1}
\end{eqnarray}

\begin{figure}
\includegraphics[width=1\linewidth]{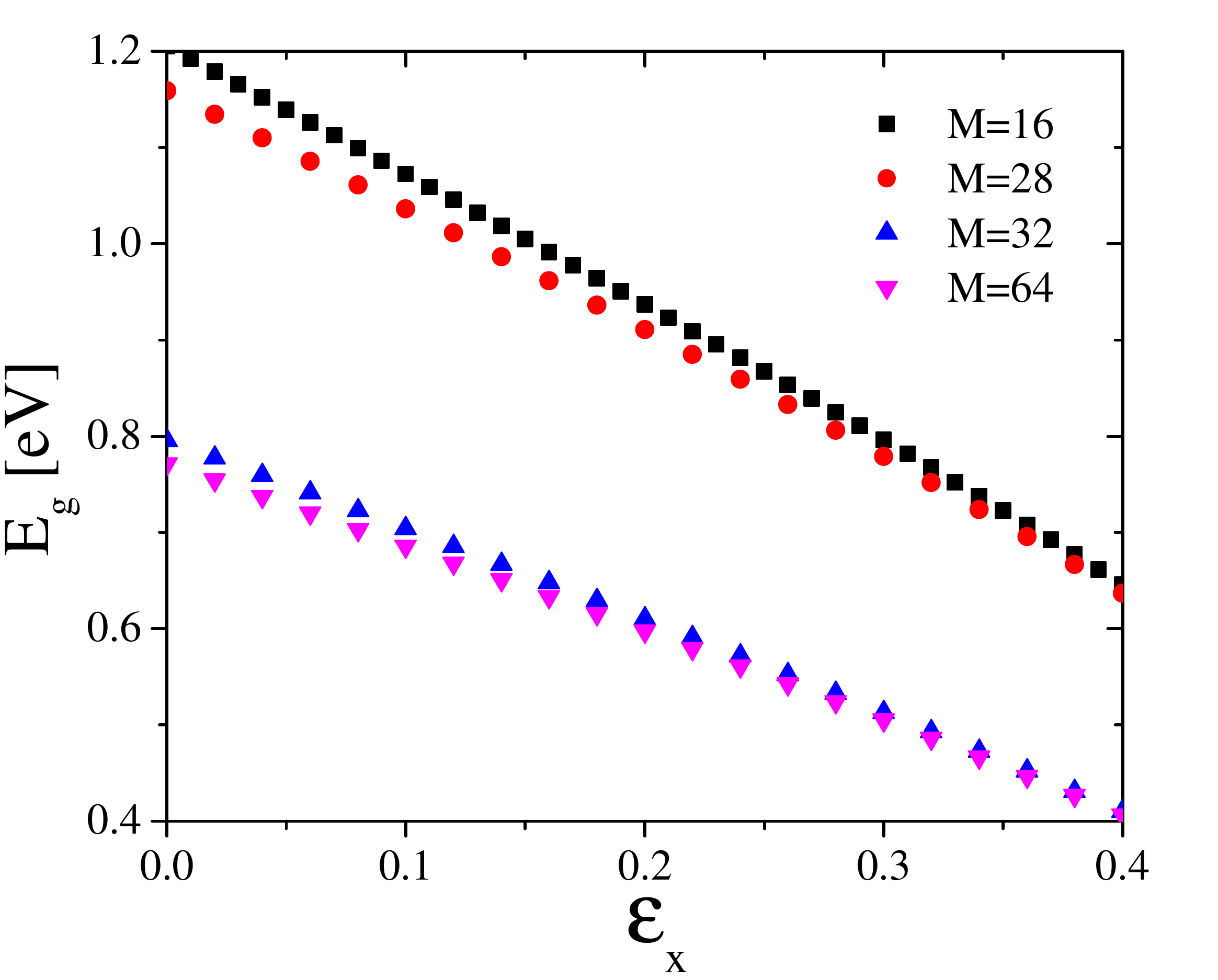}
\caption{(Color online)
The effect of strain on the electronic band gap of ZBSNRs with different widths.}
\label{fig8}
\end{figure}

We further examined dependence of band gaps of ZBSNRs on the uniaxial strain and ribbons width (see Fig.\ref{fig8}).
From the evaluated electronic band-structures, we find that the magnitude of the bandgap in each type of the zigzag B$_2$S nanoribbon is reduced as the applied strain increases. Instead, the bandgap is a linear function of the applied strain. This is consistent with previous first principles calculations of graphene nanoribbons and other 2D layered materials~\cite{Y.Lu2010,Y.Zhang12}.

It is noteworthy to mention that to further explore the width dependence of the band structures and magnetic couplings, the ZBSNRs are classified into two different family structures, the $M=4p$ and $M=4p+2$, with $p$ as an integer number.

\begin{figure}
\includegraphics[width=1\linewidth]{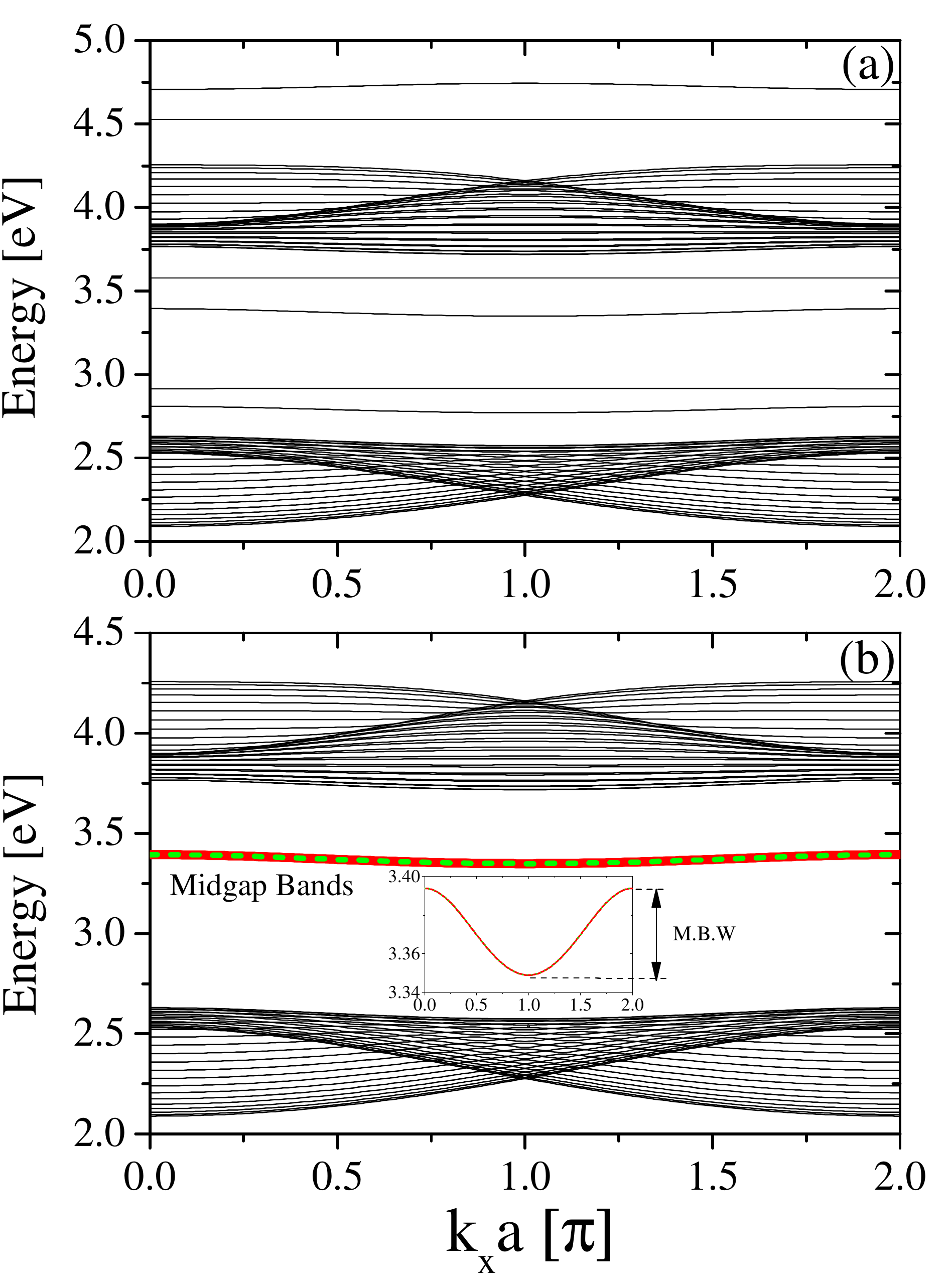}
\caption{(Color online)
The energy band structures of the undoped ($E_F=0$) infinite length nanoribbons with $M=62$ (a) and $M=64$ (b). Inset curve shows a zoomed
view of the near-midgap bands (green and red curves) with their bandwidth (BW).}
\label{DisK6264}
\end{figure}

The associated band structures of zigzag B$_2$S nanoribbons with width of $M=62$ (an example for the group $M=4p+2$) and $M=64$ (an example for the group $M=4p$), are plotted in Figs. \ref{DisK6264}(a) and \ref{DisK6264}(b), respectively.
Interestingly, for the ZBSNRs belong the $M=4p$ group, one can see that a band gap in which a pair of near-midgap bands (green and red curves) completely detached from the bulk bands, is always observed. As is known, these near-midgap energy bands are due to the edge states whose wave functions are confined near the ZBSNR edges~\cite{H.Zhang,A.CarvalhoEPL,H.Guo14}.
As shown in Fig. \ref{DisK6264}, for ZBSNRs with width of $M=4p$ only one pair of near-midgap edge modes is formed but for ZBSNRs with width of $M=4p+2$, in addition to edge states (near-midgap bands), there are other detached bulk states in the energy gap (see Fig. \ref{DisK6264}(a)).

\begin{figure}
\includegraphics[width=0.9\linewidth]{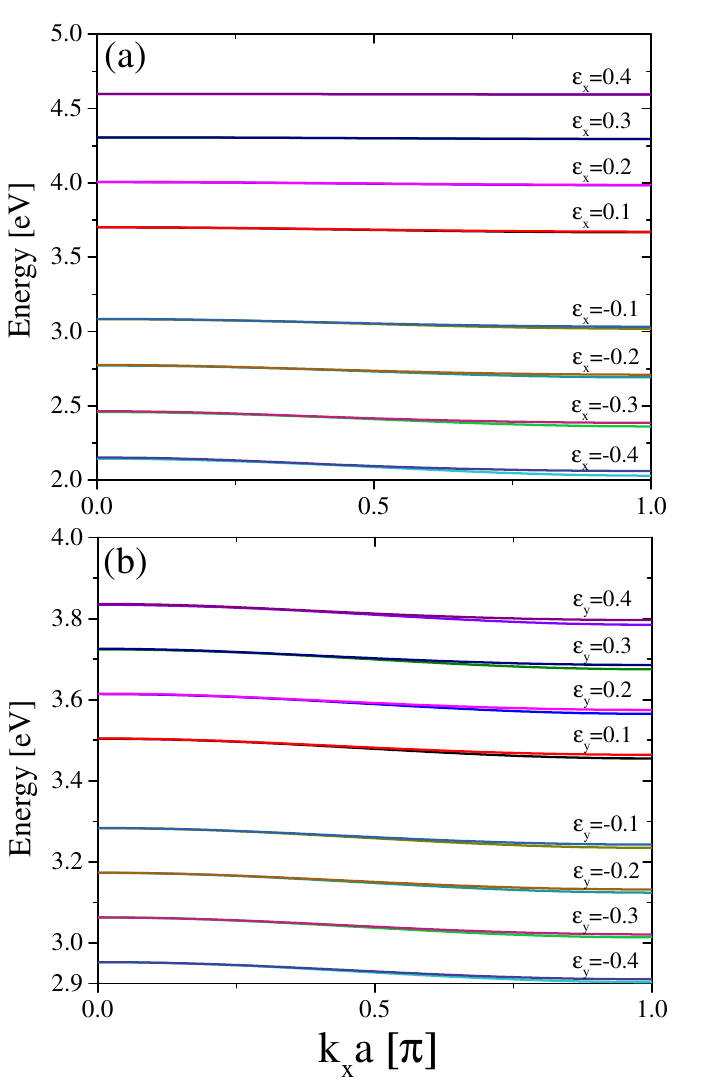}
\caption{(Color online)
Shift of the midgap energy bands under both positive and negative uniaxial strains $\varepsilon_x$ (a) and $\varepsilon_y$ (b)
Local density of states of ZBSNRs (a,b) for edge lattices and (c,d) for bulk lattices. (a,c) are for ZBSNRs with $N=150, M=18$ (example for the group $M=4p+2$), while (b,d) are for ZBSNRs with $N=150, M=16$ (example for the group $M=4p$). }
\label{figmidgapshift}
\end{figure}

It seems promising for zigzag nanoribbons to modulate the midgap energy bands, for the next generation of semiconductor devices~\cite{H.Zhang,A.CarvalhoEPL,H.Guo14}.
Thus, it is interesting to look at the bandgap and midgap states modulation in ZBSNRs. As shown in Fig.\ref{figmidgapshift}, as a key feature, the position of the midgap bands in the energy diagram of ZPNRs can be shifted by applying the in-plane strains $\varepsilon_x$ and $\varepsilon_y$.
Particularly, the near-midgap energies move up under positive strains, while shift down with negative strains.
Moreover, the near-midgap bandwidth (MBW) monotonically decreases with increasing strength of the strain and increases with the width of the ZBSNR. It is worthwhile to note that the bandwidth is generally defined as the energy difference between the upper and lower band edges.
Degeneracy of the near-midgap bands, which are always degenerate at $k_x a=0$, is increases dramatically with increasing the width of the ZPNRs.
As suggested by Soleimanikahnoj {\it et al.}, these tunability of near-midgap bands in ZBSNRs can pave the way for their potential application in pseudospin electronics based on nanoribbons~\cite{Soleimanikahnoj}.

\begin{figure}
\includegraphics[width=01\linewidth]{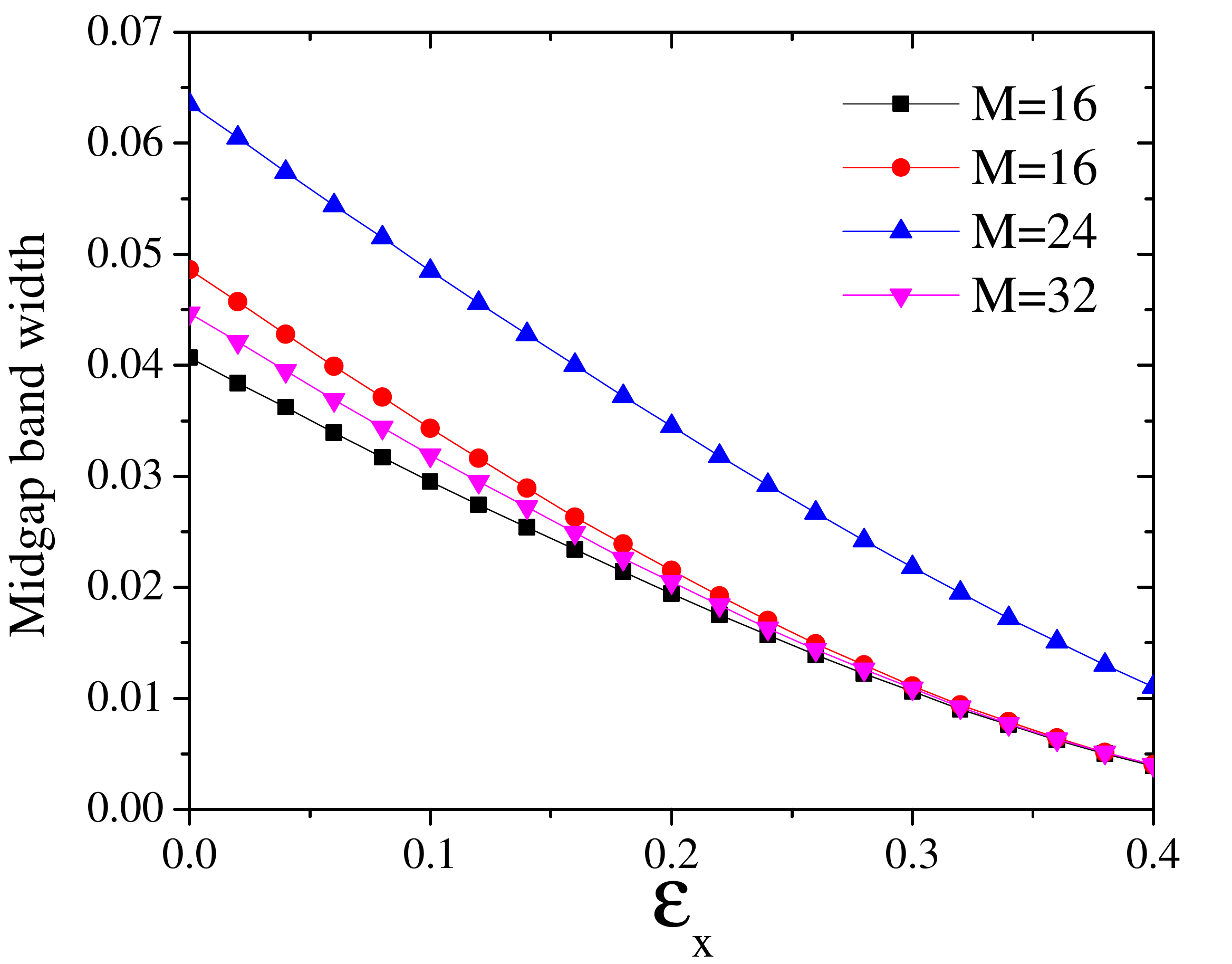}
\caption{(Color online)
The effect of strain $\varepsilon_x$ on the near-midgap bandwidth (MBW) for ZBSNRs with different widths.}
\label{FigMidg}
\end{figure}

One of the fascinating properties of the new families of 2D layered materials is their possibility to use a staggered potential to manipulate their electronic properties. Motivated by this important problem, we also examine the effect of staggered sublattice potential on the electronic structure, by breaking the discrete sublattice symmetry of this honeycomb structure. Here, we investigate the band dispersion of the ZBSNRs of infinite
length $L$ $(N \rightarrow \infty )$ under the influence of the staggered potential with various potential strength.

\begin{figure*}
\includegraphics[width=1\linewidth]{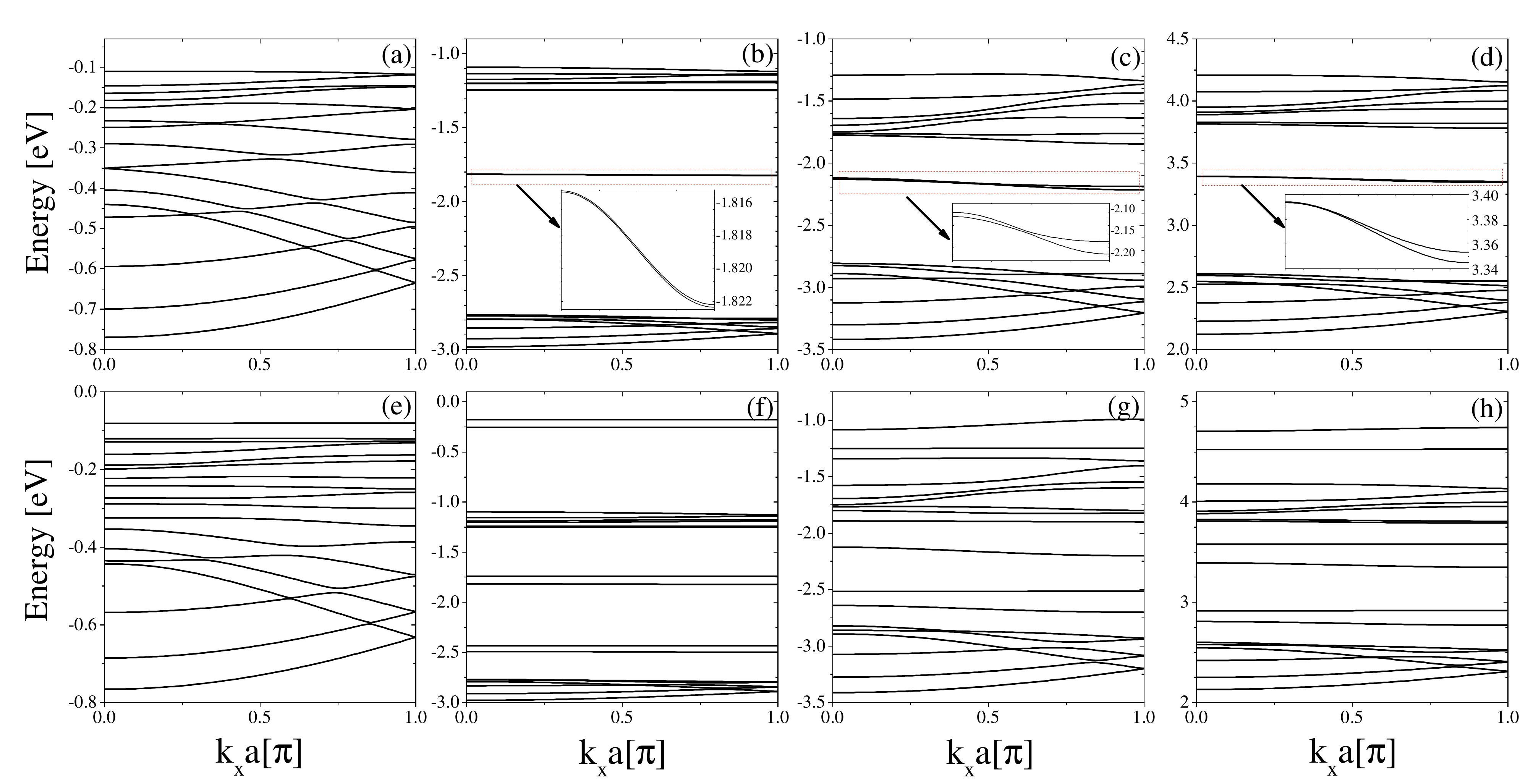}
\caption{(Color online)
The energy band structures of the infinitesized monolayer of zigzag B$_2$S nanoribbon as a function of $k_x a$ with various staggered potential parameters
(a,e) ‎$U_{S}=0, U_{B}=5.4$, (b,f) ‎$U_{S}=6.4, U_{B}=0$, (c,g) ‎$U_{S}=0, U_{B}=0$, (d,h) ‎$U_{S}=6.4, U_{B}=5.4$, all in units of eV. The top panels are for M=16 and the bottom ones are for M=18.}
\label{fig1}
\end{figure*}
The calculated band structures of zigzag ribbons of B$_2$S are shown in Fig. \ref{fig1}, for two different ribbon widths ($M=16, 18$). The top panels (a-d) are for $M=16$ and the bottom ones (e-h) are for $M=18$. The staggered sublattice potentials are as ‎$U_{S}=0, U_{B}=5.4$ (a,e), ‎$U_{S}=6.4, U_{B}=0$ (b,f), ‎$U_{S}=0, U_{B}=0$ (c,g) and ‎$U_{S}=6.4, U_{B}=5.4$  (d,h), all in units of eV. As can be seen, the resulting band structures are completely different at various values of the strength of the staggered potential.

\begin{figure}
\includegraphics[width=01\linewidth]{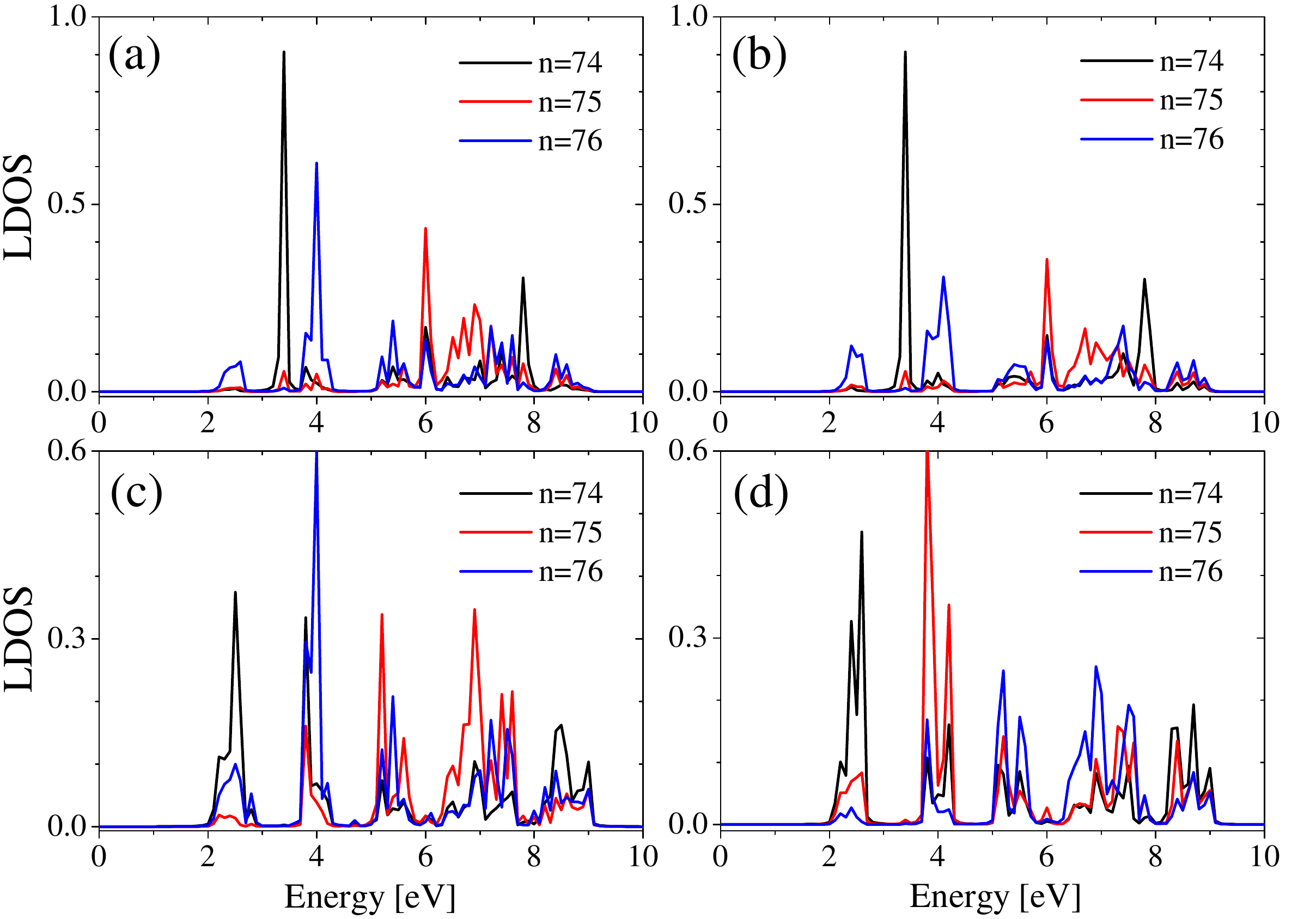}
\caption{(Color online)
Local density of states of ZBSNRs (a,b) for edge lattices and (c,d) for bulk lattices. (a,c) are for ZBSNRs with $N=150, M=18$ (example for the group $M=4p+2$), while (b,d) are for ZBSNRs with $N=150, M=16$ (example for the group $M=4p$). }
\label{fig3}
\end{figure}

As is known, understanding the sublattice-dependent of local density of states (LDOS) is essential to assess the configuration-dependent magnetic interaction.
Fig.~\ref{fig3} illustrates the LDOS of the ZBSNRs, (a,b) for edge lattices and (c,d) for bulk lattices, in which (a,c) are for ZBSNRs with $N=150$, $M=18$, while (b,d) are for ZBSNRs with $N=150$, $M=16$.
As shown, the LDOSs are effectively modulated in the ZBSNRs, for the three groups of $n=3p$, $n=3p+1$ and $n=3p+2$, where $p$ is an integer.
In Fig.~\ref{fig3}, looking at the LDOS corresponding to a various lattice sites, it is clear that the higher values of LDOS appear on the edges.
\subsection{The RKKY interaction}\label{sec:RKKY}
We are here focusing on the indirect exchange interaction in well-defined pairs of magnetic impurities with different configurations and distances on a B$_2$S surface. The carrier-mediated exchange coupling between the spin of itinerant electrons and two magnetic impurities located at positions ${\bf r}$ and ${\bf r'}$, with magnetic moments ${\bf S}_1$ and ${\bf S}_2$, is given by

\begin{equation}
V = - \lambda \ ( {\bf S}_1 \cdot {\bf s(r)} + {\bf S}_2 \cdot {\bf s(r')} ),
\label{hamilint}
\end{equation}
where ${\bf s(r)},{\bf s(r')} $ are the conduction electron spin densities at positions ${\bf r}$ and ${\bf r'}$ and $\lambda$ is the coupling between the impurity spins and the itinerant carriers.

In the linear response regime, the interaction energy between the two localized magnetic moments, derived from second-order perturbation theory, may be written as a Heisenberg form‎~\cite{Ruderman,Kasuya,Yosida,Imamura}
\begin{equation}
E({\bf r},{\bf r'}) = J ({\bf r},{\bf r'}) {\bf S}_1 \cdot {\bf S}_2,
\label{RKKYE}
\end{equation}

The RKKY interaction $J ({\bf r},{\bf r'}) $ is explained using the susceptibility, the response of the charge density $n$ to a perturbing potential $V$,
\begin{equation}
J ({\bf r},{\bf r'}) = \frac{\lambda^2 \hbar^2 }{4} \chi ({\bf r},{\bf r'}).
\label{RKKYJ}
\end{equation}

where $ \chi({\bf r},{\bf r'}) \equiv  \delta n({\bf r}) / \delta V({\bf r'})$ is the charge susceptibility for a crystal, $\delta V(r')$ is a spin-independent perturbing potential and $\delta n({\bf r})$ is the induced charge density.

The static spin susceptibility can be written in terms of the integral over the unperturbed Green's function
\begin{equation}
\chi ({\bf r},{\bf r'}) =
- \frac{2}{\pi} \int^{\varepsilon_F}_{-\infty} d\varepsilon \
{\rm Im} [G^0 ({\bf r}, {\bf r'}, \varepsilon) G^0 ({\bf r'},{\bf r}, \varepsilon)],
\label{chiGG}
\end{equation}
where $\varepsilon_F$ is the Fermi energy. The expression for the susceptibility may be obtained by using the spectral representation of the Green's function
\begin{equation}
G^0 ({\bf r},{\bf r'},\varepsilon)= \sum_{n,s} \frac{\psi_{n,s}({\bf r})\psi^{*}_{n,s}({\bf r'})}{\varepsilon+i\eta - \varepsilon_{n,s}},
\label{GFspct}
\end{equation}
where $\psi_{n,s}$ is the sublattice component of the unperturbed eigenfunction with the corresponding energy $\varepsilon_{n,s}$. For a crystalline structure, ${n,s}$ denotes the band index and spin. In other words, it just denotes a complete set of quantum states. Substituting Eq. (\ref{GFspct}) into Eq. (\ref{chiGG}), after integration over energy, we will get the result for the RKKY interaction.

The analytical background of this approach has been already described in details in previous works~\cite{moslem-si,Moslem18} and will not be rediscussed here. We only extract from these previous theoretical considerations the following desired result

\begin{eqnarray}
\chi({\bf r},{\bf r'}) &&=2 \sum_{\substack{n,,s \\ n',s'}}[ \frac{f(\varepsilon_{n,s})-f(\varepsilon_{n',s'})}{\varepsilon_{n,s}-\varepsilon_{n',s'}}\nonumber\\
&&\times \psi_{n,s}({\bf r})\psi^{* }_{n,s}({\bf r'})\psi_{n',s'}({\bf r'})\psi^{*}_{{ n'}s'}({\bf r})].
\label{chiE}
\end{eqnarray}
where, $f(\varepsilon)$, is the Fermi function.
This is a well-known formula in the linear response theory that is the main equation in this work.

\subsection{Numerical results for the RKKY interaction in zigzag B$_2$S nanoribbons }\label{sec:Numer-R}.

Here we numerically calculate the RKKY exchange interaction (Eq.\ref{chiE}) for the zigzag B$_2$S nanoribbons, based on the tight-binding model (equation ‎\ref{eqn:TBhamiltonian}‎). Note that for simplicity, all obtained data for the RKKY interaction are multiplied by $10^3$.

\begin{figure*}
\includegraphics[width=1\linewidth]{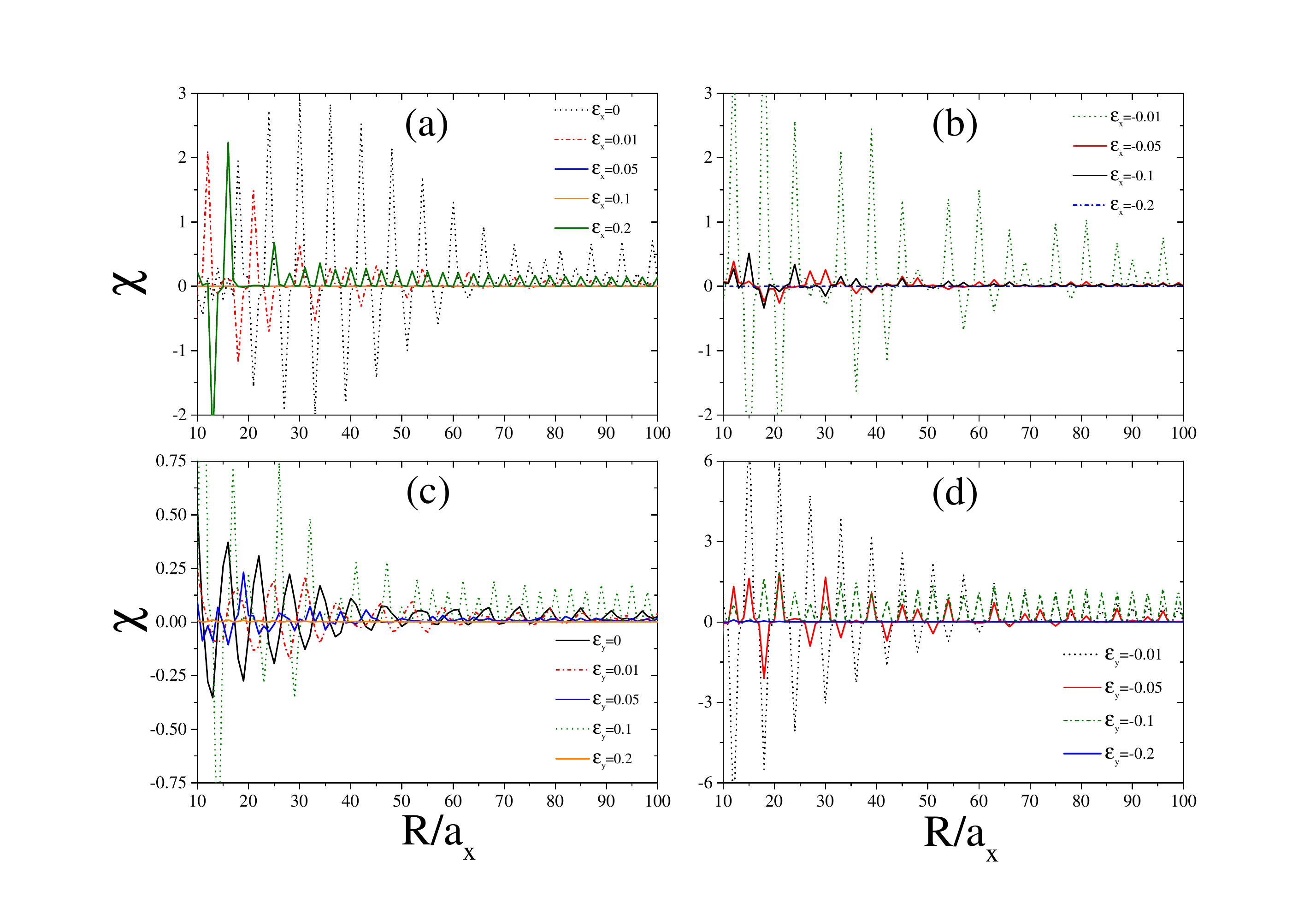}
\caption{(Color online) The variation of $\chi$ versus distance between two impurities, for doped ZBSNRs with $N=300, M=16$  for different strain strengths, $\varepsilon_x$ and $\varepsilon_y$. The Fermi energy is fixed at $E_F=2$ eV in all curves.
(a,b) When both the impurities are located on the edge, such that the one of the spins is fixed at the edge site with $(10,1)$ and another can be located on lattice points $(n,1)$. (c,d) When both the impurities are located in the interior of the ZBSNR, such that the one of spins is fixed at the lattice site with coordinate $(10,8)$ and another can be located on lattice points $(n,8)$.}
\label{fig4}
\end{figure*}

Figure \ref{fig4} shows the effective exchange interaction for doped ZBSNRs ($E_F=2$ eV ) with $N=300, M=16$, as a function of distance between the impurities for different strain strengths $\varepsilon_x$ and $\varepsilon_y$. In panels (a), (b) both the impurities are placed at the same edge, such that the one of the spins is fixed at the edge site with $(10,1)$ and another can be located on lattice points with $(n,1)$, while in panels (c), (d) both the impurities are located in the interior of the ZBSNR, such that the one of the spins is fixed at the lattice site with coordinate $(10,8)$ and another can be located on lattice points $(n,8)$.

\begin{figure*}
\includegraphics[width=1\linewidth]{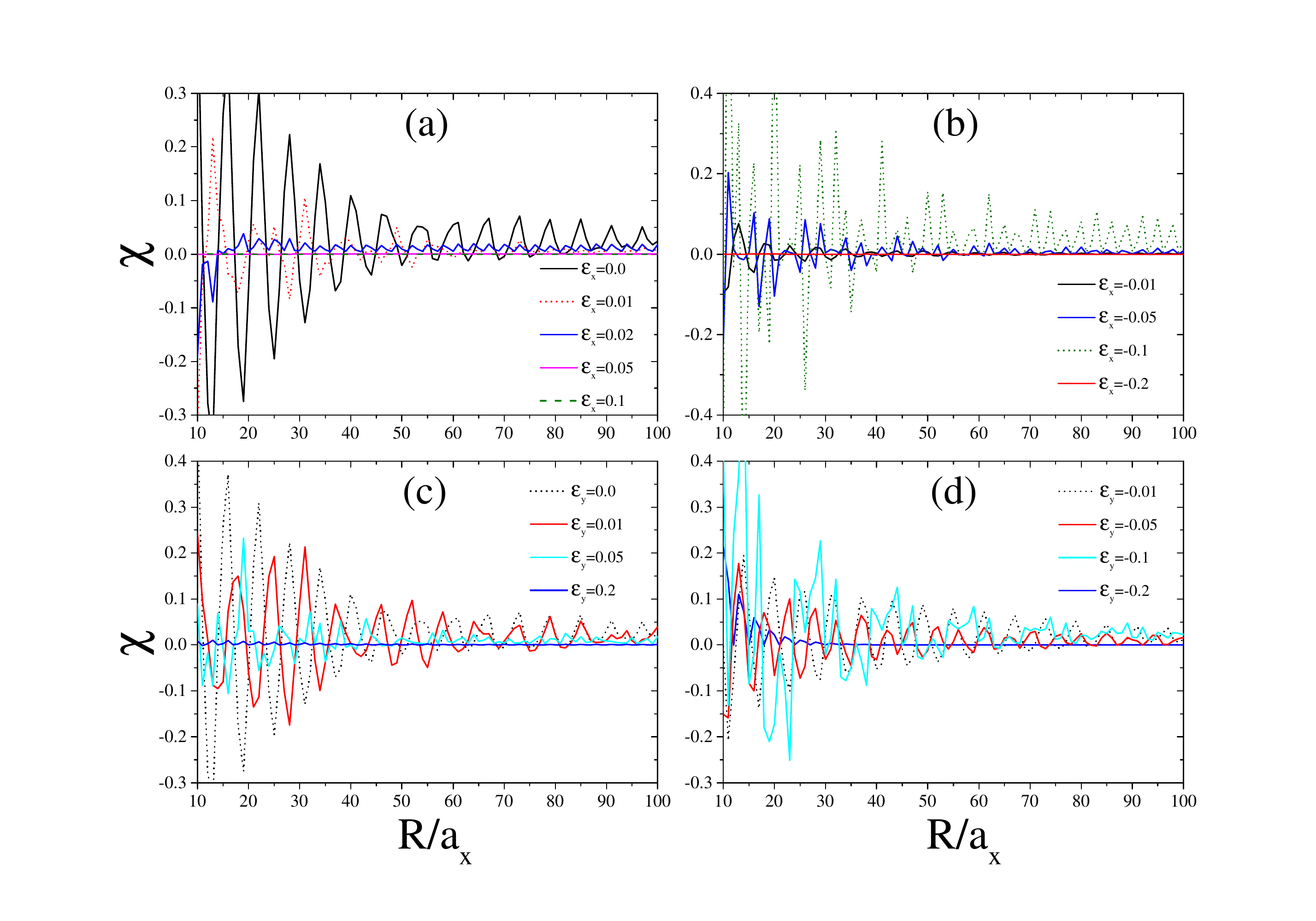}
\caption{(Color online) The variation of $\chi$ versus distance between two impurities, for doped ZBSNRs with $N=300, M=18$  for different strain strengths, $\varepsilon_x$ and $\varepsilon_y$. The Fermi energy is fixed at $E_F=2$ eV in all curves.
(a,b) When both the impurities are located on the edge, such that the one of the spins is fixed at the edge site with $(10,1)$ and another can be located on lattice points $(n,1)$. (c,d) When both the impurities are located in the interior of the ZBSNR, such that the one of spins is fixed at the lattice site with coordinate $(10,8)$ and another can be located on lattice points $(n,8)$.}
\label{fig5}
\end{figure*}

What the Fig. \ref{fig5} shows is the same as Fig. \ref{fig4} but for ZBSNRs with $N=300, M=18$,
In the panels (a), (b) both the impurities are located on the same edge, such that the one of the spins is fixed at the edge site with $(10,1)$ and another can be located on lattice points $(n,1)$ and in panels (c), (d) both the impurities are located in the interior of the ZBSNR, such that the one of the spins is fixed at the lattice site with coordinate $(10,8)$ and another can be located on lattice points $(n,8)$.

\begin{figure}
\includegraphics[width=1.0\linewidth]{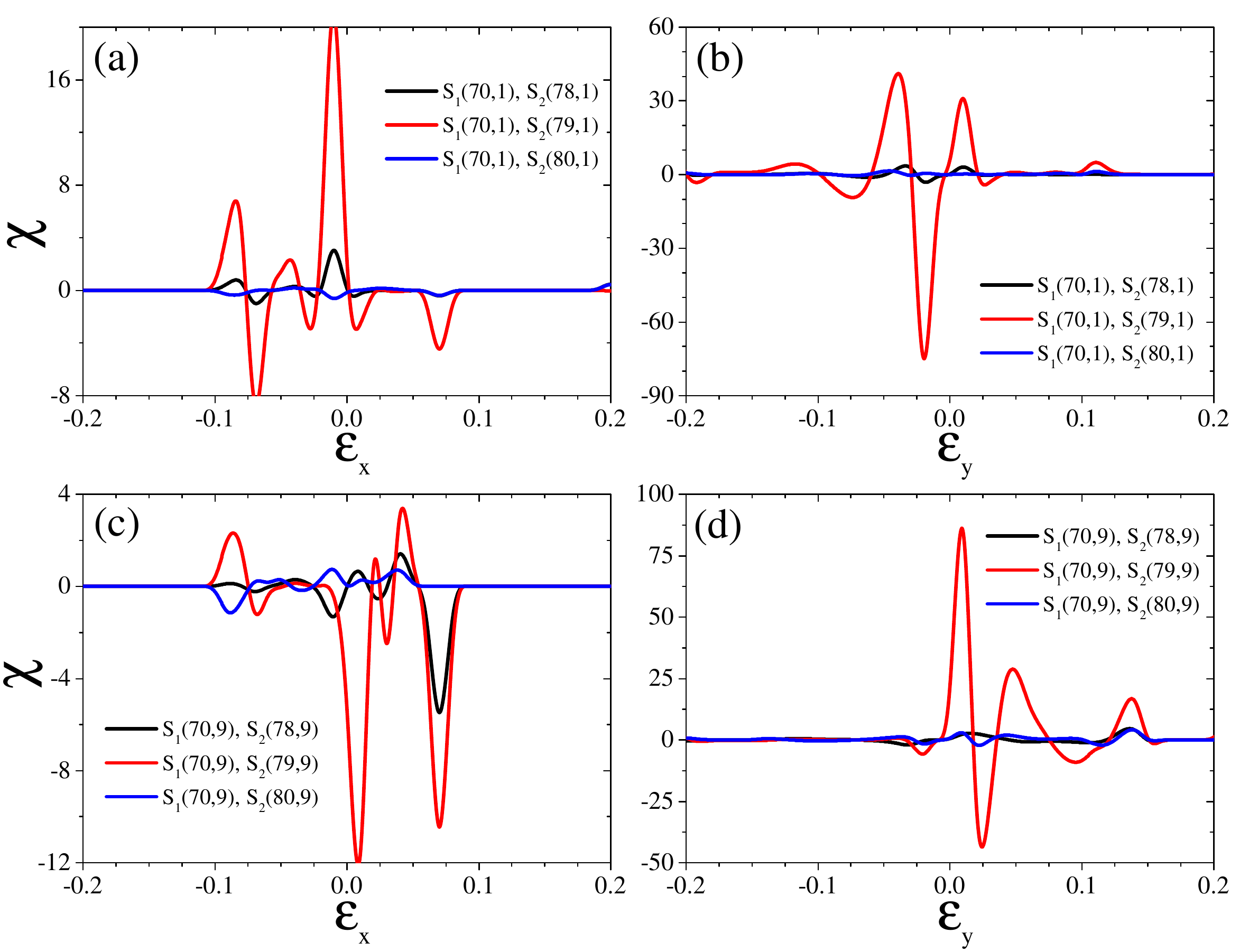}
\caption{(Color online) The variation of $\chi$ with strain $\varepsilon_x$ (a,c) and  $\varepsilon_y$ (b,d), for different distance configurations, for $N=150 , M=18$ (the $M=4p+2$ family of ZBSNRs). Top panels (a,b) are for two impurities placed on the same edge and bottom panels(c,d) are for both the impurities are within the bulk of ZBSNR.}
\label{fig6}
\end{figure}

The strain engineering of the magnetic ground state in dilute semiconductors and 2D materials is a perfect platform for their implementation in nano devices. To gain further insight into this practical mechanism, the role of strain on the magnetic interaction of ZBSNRs is also explored. To this end, we have shown that the magnetic properties of the ZBSNRs can be effectively tuned by mechanical strains, as shown in Figs.~\ref{fig6} and \ref{fig7}.
In the panels (a,b) both the impurities are located on the same edge, such that the one of the spins is fixed at the edge site with coordinate $(10,1)$ and another can be located on lattice points $(n,1)$, while in panels (c,d) both the impurities are located within the bulk of ZBSNR, such that the one of spins is fixed at the lattice site with coordinate $(10,8)$ and another can be located on lattice points $(n,8)$.

It is shown that the RKKY coupling between adsorbed impurities at a fixed distance on the ZBSNR has an oscillatory behaviour versus strains, due to the strain dependence of the Fermi wave vector.
The sign-changing oscillations of the exchange coupling, which appear in terms of strains $\varepsilon_x$ and $\varepsilon_y$, are very interesting and may have significant implications for strain engineering of the magnetic ground state in ZBSNRs.

The quenching of the short-range RKKY interaction at and below a certain strain is seen in these figures. It is clear that the distance configuration of the magnetic impurities has a very significant impact on the strain engineering of magnetic coupling in 2D ZBSNRs. For instance for ZBSNRs with $M=18$ (a $M=4p+2$ ZBSNR family), whether the impurities are on the edge or inside the ZBSNR, when the first moment is pinned on a $n=3p$ family site (here $n=69$), if the second one is seated on a $n=3p+1$ family site (here $n=79$), the observed peaks of the RKKY oscillations are stronger.

For the case of a ZBSNR of width $M=20$ (a $M=4p$ family) (see Fig.\label{fig7}), in the case when both the impurities are placed at the same edge, if the first moment is pinned at a $n=3p$ family site (here $n=69$), when the second one is seated on a $n=3p$ family site (here $n=79$), the observed peaks of the RKKY oscillations are stronger. But in the case when both the impurities are away from the edge (when both the impurities are inside the bulk, along the line $m=10$), the strongest peaks are related to the $n=3p$ family site (here $n=78$).

\begin{figure}
\includegraphics[width=1.0\linewidth]{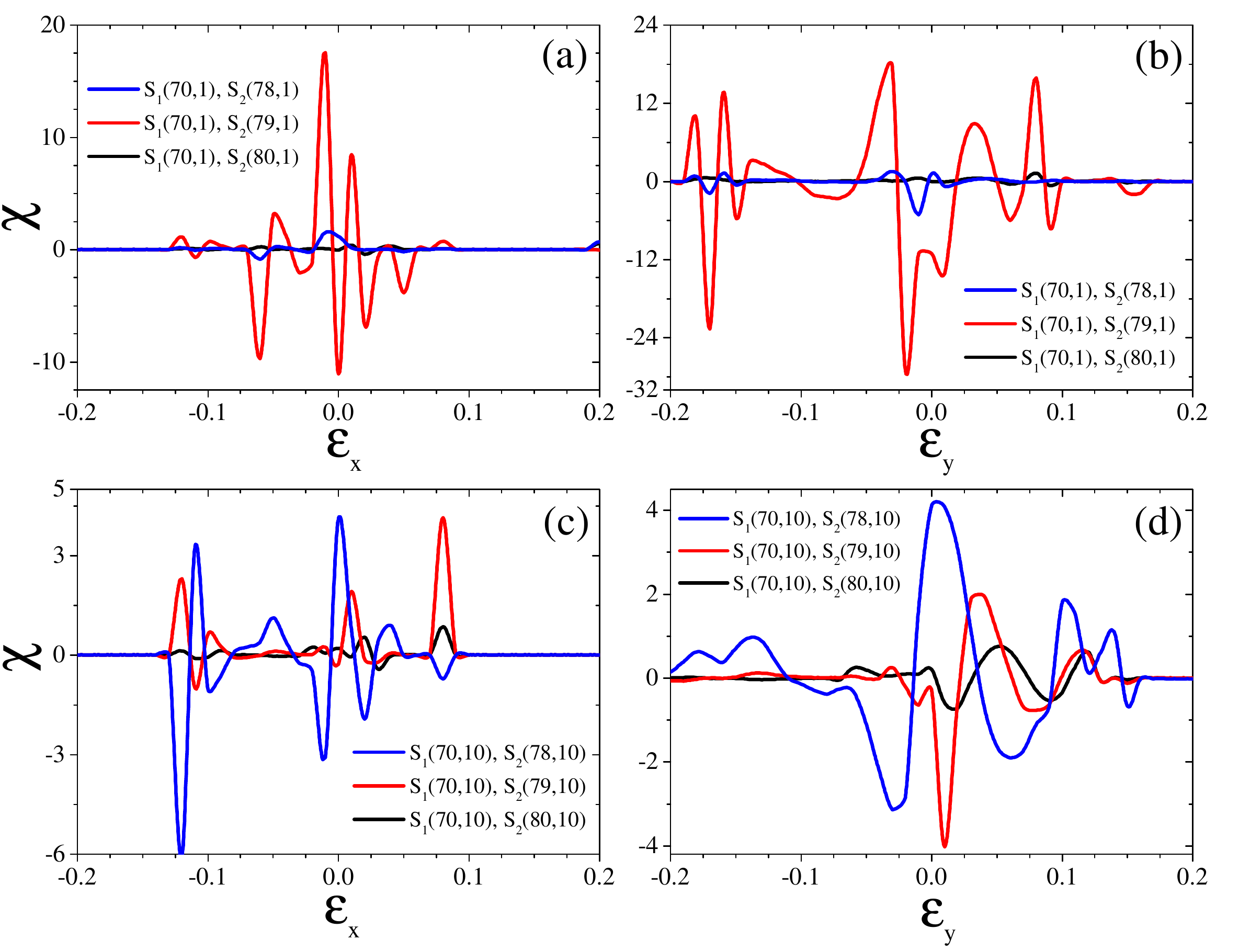}
\caption{(Color online) The variation of $\chi$ with strain $\varepsilon_x$ (a,c) and  $\varepsilon_y$ (b,d), for different distance configurations, for $N=150 , M=20$ (the $M=4p$ family of ZBSNRs), (a,b) for two impurities placed on the same edge and (c,d) for both the impurities are within the bulk of ZBSNR.}
\label{fig7}
\end{figure}

\section{summary}\label{sec:summary}.
To summarize, in this work, we numerically investigate the RKKY exchange coupling between two magnetic impurities located on a zigzag B$_2$S nanoribbon, a recently realized monolayer, as a strained graphene.
In the first part of our study, we focus on how the electronic structure of monolayer nanoribbons of B$_2$S is modified under uniaxial strains, then  employing a tight-binding approach together with the conventional theory of elasticity, we discuss how strain-induced local deformations can be used as a means to affect Ruderman-Kittel-Kasuya-Yosida (RKKY) interaction in zigzag nanoribbons of B$_2$S.
We show that breaking inversion symmetry in zigzag B$_2$S nanoribbons (ZBSNR), e.g., by introducing staggered sublattice potentials, play a key role in the modulation of their electronic properties.
More interestingly, for the ZBSNRs belong to the group $M=4p$, with $M$ the width of the ZBSNR and $p$ an integer number, one can see that a band gap, in which a pair of near-midgap bands completely detached from the bulk bands is always observed.
As a key feature, the position of the midgap bands in the energy diagram of ZBSNRs can be shifted by applying the in-plane strains $\varepsilon_x$ and $\varepsilon_y$. Moreover, the near-midgap bandwidth (MBW) monotonically decreases with increasing strength of the strain and increases with the width of the ZBSNR. The energy gap of the ZBSNRs decreased with increasing the applied strain and ribbon width.
The spatial and strain dependency of the exchange interaction in various configurations of the magnetic impurities are also evaluated.
It is shown that that magnetic interactions between adsorbed magnetic impurities in B$_2$S nanoribbons can be manipulated by careful strain engineering of such systems. We found that the RKKY coupling between adsorbed impurities at a fixed distance on the ZBSNR has an oscillatory behaviour versus strains, due to the strain dependence of the Fermi wave vector.
Our results suggest that these tunable electronic and magnetic properties of ZBSNRs mean they may find applications in spintronics and pseudospin electronics based on monolayer B$_2$S nanoribbons.

\section*{ACKNOWLEDGMENT}

This work is partially supported by Iran Science Elites Federation.
\section*{References}


\begin{thebibliography}{51}%
\makeatletter
\providecommand \@ifxundefined [1]{%
 \@ifx{#1\undefined}
}%
\providecommand \@ifnum [1]{%
 \ifnum #1\expandafter \@firstoftwo
 \else \expandafter \@secondoftwo
 \fi
}%
\providecommand \@ifx [1]{%
 \ifx #1\expandafter \@firstoftwo
 \else \expandafter \@secondoftwo
 \fi
}%
\providecommand \natexlab [1]{#1}%
\providecommand \enquote  [1]{``#1''}%
\providecommand \bibnamefont  [1]{#1}%
\providecommand \bibfnamefont [1]{#1}%
\providecommand \citenamefont [1]{#1}%
\providecommand \href@noop [0]{\@secondoftwo}%
\providecommand \href [0]{\begingroup \@sanitize@url \@href}%
\providecommand \@href[1]{\@@startlink{#1}\@@href}%
\providecommand \@@href[1]{\endgroup#1\@@endlink}%
\providecommand \@sanitize@url [0]{\catcode `\\12\catcode `\$12\catcode
  `\&12\catcode `\#12\catcode `\^12\catcode `\_12\catcode `\%12\relax}%
\providecommand \@@startlink[1]{}%
\providecommand \@@endlink[0]{}%
\providecommand \url  [0]{\begingroup\@sanitize@url \@url }%
\providecommand \@url [1]{\endgroup\@href {#1}{\urlprefix }}%
\providecommand \urlprefix  [0]{URL }%
\providecommand \Eprint [0]{\href }%
\providecommand \doibase [0]{http://dx.doi.org/}%
\providecommand \selectlanguage [0]{\@gobble}%
\providecommand \bibinfo  [0]{\@secondoftwo}%
\providecommand \bibfield  [0]{\@secondoftwo}%
\providecommand \translation [1]{[#1]}%
\providecommand \BibitemOpen [0]{}%
\providecommand \bibitemStop [0]{}%
\providecommand \bibitemNoStop [0]{.\EOS\space}%
\providecommand \EOS [0]{\spacefactor3000\relax}%
\providecommand \BibitemShut  [1]{\csname bibitem#1\endcsname}%
\let\auto@bib@innerbib\@empty
\bibitem [{\citenamefont {Ruderman}\ and\ \citenamefont
  {Kittel}(1954)}]{Ruderman}%
  \BibitemOpen
  \bibfield  {author} {\bibinfo {author} {\bibfnamefont {M.~A.}\ \bibnamefont
  {Ruderman}}\ and\ \bibinfo {author} {\bibfnamefont {C.}~\bibnamefont
  {Kittel}},\ }\href {\doibase 10.1103/PhysRev.96.99} {\bibfield  {journal}
  {\bibinfo  {journal} {Phys. Rev.}\ }\textbf {\bibinfo {volume} {96}},\
  \bibinfo {pages} {99} (\bibinfo {year} {1954})}\BibitemShut {NoStop}%
\bibitem [{\citenamefont {Kasuya}(1956)}]{Kasuya}%
  \BibitemOpen
  \bibfield  {author} {\bibinfo {author} {\bibfnamefont {T.}~\bibnamefont
  {Kasuya}},\ }\href {\doibase 10.1143/PTP.16.45} {\bibfield  {journal}
  {\bibinfo  {journal} {Prog. Theor. Phys.}\ }\textbf {\bibinfo {volume}
  {16}},\ \bibinfo {pages} {45} (\bibinfo {year} {1956})}\BibitemShut {NoStop}%
\bibitem [{\citenamefont {Yosida}(1957)}]{Yosida}%
  \BibitemOpen
  \bibfield  {author} {\bibinfo {author} {\bibfnamefont {K.}~\bibnamefont
  {Yosida}},\ }\href {\doibase 10.1103/PhysRev.106.893} {\bibfield  {journal}
  {\bibinfo  {journal} {Phys. Rev.}\ }\textbf {\bibinfo {volume} {106}},\
  \bibinfo {pages} {893} (\bibinfo {year} {1957})}\BibitemShut {NoStop}%
\bibitem [{\citenamefont {Vozmediano}\ \emph {et~al.}(2005)\citenamefont
  {Vozmediano}, \citenamefont {Lopez-Sancho}, \citenamefont {Stauber},\ and\
  \citenamefont {Guinea}}]{Vozmediano}%
  \BibitemOpen
  \bibfield  {author} {\bibinfo {author} {\bibfnamefont {M.~A.~H.}\
  \bibnamefont {Vozmediano}}, \bibinfo {author} {\bibfnamefont {M.~P.}\
  \bibnamefont {Lopez-Sancho}}, \bibinfo {author} {\bibfnamefont
  {T.}~\bibnamefont {Stauber}}, \ and\ \bibinfo {author} {\bibfnamefont
  {F.}~\bibnamefont {Guinea}},\ }\href {\doibase 10.1103/PhysRevB.72.155121}
  {\bibfield  {journal} {\bibinfo  {journal} {Phys. Rev. B}\ }\textbf {\bibinfo
  {volume} {72}},\ \bibinfo {pages} {155121} (\bibinfo {year}
  {2005})}\BibitemShut {NoStop}%
\bibitem [{\citenamefont {Brey}\ \emph {et~al.}(2007)\citenamefont {Brey},
  \citenamefont {Fertig},\ and\ \citenamefont {Sarma}}]{Brey}%
  \BibitemOpen
  \bibfield  {author} {\bibinfo {author} {\bibfnamefont {L.}~\bibnamefont
  {Brey}}, \bibinfo {author} {\bibfnamefont {H.}~\bibnamefont {Fertig}}, \ and\
  \bibinfo {author} {\bibfnamefont {S.~D.}\ \bibnamefont {Sarma}},\ }\href
  {\doibase 10.1103/PhysRevLett.99.116802} {\bibfield  {journal} {\bibinfo
  {journal} {Phys. Rev. Lett.}\ }\textbf {\bibinfo {volume} {99}},\ \bibinfo
  {pages} {116802} (\bibinfo {year} {2007})}\BibitemShut {NoStop}%
\bibitem [{\citenamefont {D.J.~Priour}\ \emph {et~al.}(2004)\citenamefont
  {D.J.~Priour}, \citenamefont {Hwang},\ and\ \citenamefont {Sarma}}]{Priour}%
  \BibitemOpen
  \bibfield  {author} {\bibinfo {author} {\bibfnamefont {J.}~\bibnamefont
  {D.J.~Priour}}, \bibinfo {author} {\bibfnamefont {E.~H.}\ \bibnamefont
  {Hwang}}, \ and\ \bibinfo {author} {\bibfnamefont {S.~D.}\ \bibnamefont
  {Sarma}},\ }\href {\doibase 10.1103/PhysRevLett.92.117201} {\bibfield
  {journal} {\bibinfo  {journal} {Phys. Rev. Lett.}\ }\textbf {\bibinfo
  {volume} {92}},\ \bibinfo {pages} {117201} (\bibinfo {year}
  {2004})}\BibitemShut {NoStop}%
\bibitem [{\citenamefont {Matsukura}\ \emph {et~al.}(1998)\citenamefont
  {Matsukura}, \citenamefont {Ohno}, \citenamefont {Shen},\ and\ \citenamefont
  {Sugawara}}]{Matsukura}%
  \BibitemOpen
  \bibfield  {author} {\bibinfo {author} {\bibfnamefont {F.}~\bibnamefont
  {Matsukura}}, \bibinfo {author} {\bibfnamefont {H.}~\bibnamefont {Ohno}},
  \bibinfo {author} {\bibfnamefont {A.}~\bibnamefont {Shen}}, \ and\ \bibinfo
  {author} {\bibfnamefont {Y.}~\bibnamefont {Sugawara}},\ }\href {\doibase
  10.1103/PhysRevB.57.R2037} {\bibfield  {journal} {\bibinfo  {journal} {Phys.
  Rev. B}\ }\textbf {\bibinfo {volume} {57}},\ \bibinfo {pages} {2037(R)}
  (\bibinfo {year} {1998})}\BibitemShut {NoStop}%
\bibitem [{\citenamefont {Ko}\ \emph {et~al.}(2011)\citenamefont {Ko},
  \citenamefont {Kim}, \citenamefont {Kim}, \citenamefont {Kim}, \citenamefont
  {Kim}, \citenamefont {Min}, \citenamefont {Park}, \citenamefont {Chang},
  \citenamefont {Lin}, \citenamefont {Tanaka},\ and\ \citenamefont
  {Cheong}}]{Ko}%
  \BibitemOpen
  \bibfield  {author} {\bibinfo {author} {\bibfnamefont {K.~T.}\ \bibnamefont
  {Ko}}, \bibinfo {author} {\bibfnamefont {K.}~\bibnamefont {Kim}}, \bibinfo
  {author} {\bibfnamefont {S.~B.}\ \bibnamefont {Kim}}, \bibinfo {author}
  {\bibfnamefont {H.~D.}\ \bibnamefont {Kim}}, \bibinfo {author} {\bibfnamefont
  {J.~Y.}\ \bibnamefont {Kim}}, \bibinfo {author} {\bibfnamefont {B.~I.}\
  \bibnamefont {Min}}, \bibinfo {author} {\bibfnamefont {J.~H.}\ \bibnamefont
  {Park}}, \bibinfo {author} {\bibfnamefont {F.~H.}\ \bibnamefont {Chang}},
  \bibinfo {author} {\bibfnamefont {H.~J.}\ \bibnamefont {Lin}}, \bibinfo
  {author} {\bibfnamefont {A.}~\bibnamefont {Tanaka}}, \ and\ \bibinfo {author}
  {\bibfnamefont {S.~W.}\ \bibnamefont {Cheong}},\ }\href {\doibase
  10.1103/PhysRevLett.107.247201} {\bibfield  {journal} {\bibinfo  {journal}
  {Phys. Rev. Lett.}\ }\textbf {\bibinfo {volume} {107}},\ \bibinfo {pages}
  {247201} (\bibinfo {year} {2011})}\BibitemShut {NoStop}%
\bibitem [{\citenamefont {Ohno}(1998)}]{Ohno-science}%
  \BibitemOpen
  \bibfield  {author} {\bibinfo {author} {\bibfnamefont {H.}~\bibnamefont
  {Ohno}},\ }\href {\doibase 10.1126/science.281.5379.951} {\bibfield
  {journal} {\bibinfo  {journal} {Science}\ }\textbf {\bibinfo {volume}
  {281}},\ \bibinfo {pages} {951} (\bibinfo {year} {1998})}\BibitemShut
  {NoStop}%
\bibitem [{\citenamefont {Minamitani}\ \emph {et~al.}(2010)\citenamefont
  {Minamitani}, \citenamefont {Dino}, \citenamefont {Nakanishi},\ and\
  \citenamefont {Kasai}}]{Minamitani}%
  \BibitemOpen
  \bibfield  {author} {\bibinfo {author} {\bibfnamefont {E.}~\bibnamefont
  {Minamitani}}, \bibinfo {author} {\bibfnamefont {W.~A.}\ \bibnamefont
  {Dino}}, \bibinfo {author} {\bibfnamefont {H.}~\bibnamefont {Nakanishi}}, \
  and\ \bibinfo {author} {\bibfnamefont {H.}~\bibnamefont {Kasai}},\ }\href
  {\doibase 10.1103/PhysRevB.82.153203} {\bibfield  {journal} {\bibinfo
  {journal} {Phys. Rev. B}\ }\textbf {\bibinfo {volume} {82}},\ \bibinfo
  {pages} {153203} (\bibinfo {year} {2010})}\BibitemShut {NoStop}%
\bibitem [{\citenamefont {Hsu}\ \emph {et~al.}(2015)\citenamefont {Hsu},
  \citenamefont {Stano}, \citenamefont {Klinovaja},\ and\ \citenamefont
  {Loss}}]{Loss15}%
  \BibitemOpen
  \bibfield  {author} {\bibinfo {author} {\bibfnamefont {C.-H.}\ \bibnamefont
  {Hsu}}, \bibinfo {author} {\bibfnamefont {P.}~\bibnamefont {Stano}}, \bibinfo
  {author} {\bibfnamefont {J.}~\bibnamefont {Klinovaja}}, \ and\ \bibinfo
  {author} {\bibfnamefont {D.}~\bibnamefont {Loss}},\ }\href {\doibase
  10.1103/PhysRevB.92.235435} {\bibfield  {journal} {\bibinfo  {journal} {Phys.
  Rev. B}\ }\textbf {\bibinfo {volume} {92}},\ \bibinfo {pages} {235435}
  (\bibinfo {year} {2015})}\BibitemShut {NoStop}%
\bibitem [{\citenamefont {Zare}\ \emph {et~al.}(2016)\citenamefont {Zare},
  \citenamefont {Parhizgar},\ and\ \citenamefont {Asgari}}]{moslem-si}%
  \BibitemOpen
  \bibfield  {author} {\bibinfo {author} {\bibfnamefont {M.}~\bibnamefont
  {Zare}}, \bibinfo {author} {\bibfnamefont {F.}~\bibnamefont {Parhizgar}}, \
  and\ \bibinfo {author} {\bibfnamefont {R.}~\bibnamefont {Asgari}},\ }\href
  {\doibase 10.1103/PhysRevB.94.045443} {\bibfield  {journal} {\bibinfo
  {journal} {Phys. Rev. B}\ }\textbf {\bibinfo {volume} {94}},\ \bibinfo
  {pages} {045443} (\bibinfo {year} {2016})}\BibitemShut {NoStop}%
\bibitem [{\citenamefont {Shiranzaei}\ \emph
  {et~al.}(2017{\natexlab{a}})\citenamefont {Shiranzaei}, \citenamefont
  {Cheraghchi},\ and\ \citenamefont {Parhizgar}}]{Mahroo-RKKY}%
  \BibitemOpen
  \bibfield  {author} {\bibinfo {author} {\bibfnamefont {M.}~\bibnamefont
  {Shiranzaei}}, \bibinfo {author} {\bibfnamefont {H.}~\bibnamefont
  {Cheraghchi}}, \ and\ \bibinfo {author} {\bibfnamefont {F.}~\bibnamefont
  {Parhizgar}},\ }\href {\doibase 10.1103/PhysRevB.96.024413} {\bibfield
  {journal} {\bibinfo  {journal} {Phys. Rev. B}\ }\textbf {\bibinfo {volume}
  {96}},\ \bibinfo {pages} {024413} (\bibinfo {year}
  {2017}{\natexlab{a}})}\BibitemShut {NoStop}%
\bibitem [{\citenamefont {Abanin}\ and\ \citenamefont {Pesin}(2011)}]{pesin}%
  \BibitemOpen
  \bibfield  {author} {\bibinfo {author} {\bibfnamefont {D.~A.}\ \bibnamefont
  {Abanin}}\ and\ \bibinfo {author} {\bibfnamefont {D.~A.}\ \bibnamefont
  {Pesin}},\ }\href {\doibase 10.1103/PhysRevLett.106.136802} {\bibfield
  {journal} {\bibinfo  {journal} {Phys. Rev. Lett.}\ }\textbf {\bibinfo
  {volume} {106}},\ \bibinfo {pages} {136802} (\bibinfo {year}
  {2011})}\BibitemShut {NoStop}%
\bibitem [{\citenamefont {Eggenkamp}\ \emph {et~al.}(1995)\citenamefont
  {Eggenkamp}, \citenamefont {Swagten}, \citenamefont {Story}, \citenamefont
  {Litvinov}, \citenamefont {Swüste},\ and\ \citenamefont
  {de~Jonge}}]{Eggenkamp}%
  \BibitemOpen
  \bibfield  {author} {\bibinfo {author} {\bibfnamefont {P.~J.~T.}\
  \bibnamefont {Eggenkamp}}, \bibinfo {author} {\bibfnamefont {H.~J.~M.}\
  \bibnamefont {Swagten}}, \bibinfo {author} {\bibfnamefont {T.}~\bibnamefont
  {Story}}, \bibinfo {author} {\bibfnamefont {V.~I.}\ \bibnamefont {Litvinov}},
  \bibinfo {author} {\bibfnamefont {C.~H.~W.}\ \bibnamefont {Swüste}}, \ and\
  \bibinfo {author} {\bibfnamefont {W.~J.~M.}\ \bibnamefont {de~Jonge}},\
  }\href {\doibase 10.1103/PhysRevB.51.15250} {\bibfield  {journal} {\bibinfo
  {journal} {Phys. Rev. B}\ }\textbf {\bibinfo {volume} {51}},\ \bibinfo
  {pages} {15250} (\bibinfo {year} {1995})}\BibitemShut {NoStop}%
\bibitem [{\citenamefont {sui Liu}\ \emph {et~al.}(1987)\citenamefont {sui
  Liu}, \citenamefont {Roshen},\ and\ \citenamefont {Ruvalds}}]{Liu87}%
  \BibitemOpen
  \bibfield  {author} {\bibinfo {author} {\bibfnamefont {F.}~\bibnamefont {sui
  Liu}}, \bibinfo {author} {\bibfnamefont {W.~A.}\ \bibnamefont {Roshen}}, \
  and\ \bibinfo {author} {\bibfnamefont {J.}~\bibnamefont {Ruvalds}},\ }\href
  {\doibase 10.1103/PhysRevB.36.492} {\bibfield  {journal} {\bibinfo  {journal}
  {Phys. Rev. B}\ }\textbf {\bibinfo {volume} {36}},\ \bibinfo {pages} {492}
  (\bibinfo {year} {1987})}\BibitemShut {NoStop}%
\bibitem [{\citenamefont {Parhizgar}\ \emph
  {et~al.}(2013{\natexlab{a}})\citenamefont {Parhizgar}, \citenamefont
  {Sherafati}, \citenamefont {Asgari},\ and\ \citenamefont
  {Satpathy}}]{fariborz-blg}%
  \BibitemOpen
  \bibfield  {author} {\bibinfo {author} {\bibfnamefont {F.}~\bibnamefont
  {Parhizgar}}, \bibinfo {author} {\bibfnamefont {M.}~\bibnamefont
  {Sherafati}}, \bibinfo {author} {\bibfnamefont {R.}~\bibnamefont {Asgari}}, \
  and\ \bibinfo {author} {\bibfnamefont {S.}~\bibnamefont {Satpathy}},\ }\href
  {\doibase 10.1103/PhysRevB.87.165429} {\bibfield  {journal} {\bibinfo
  {journal} {Phys. Rev. B}\ }\textbf {\bibinfo {volume} {16}},\ \bibinfo
  {pages} {165429} (\bibinfo {year} {2013}{\natexlab{a}})}\BibitemShut
  {NoStop}%
\bibitem [{\citenamefont {Parhizgar}\ \emph
  {et~al.}(2013{\natexlab{b}})\citenamefont {Parhizgar}, \citenamefont
  {Rostami},\ and\ \citenamefont {Asgari}}]{fariborz-mos2}%
  \BibitemOpen
  \bibfield  {author} {\bibinfo {author} {\bibfnamefont {F.}~\bibnamefont
  {Parhizgar}}, \bibinfo {author} {\bibfnamefont {H.}~\bibnamefont {Rostami}},
  \ and\ \bibinfo {author} {\bibfnamefont {R.}~\bibnamefont {Asgari}},\ }\href
  {\doibase 10.1103/PhysRevB.87.125401} {\bibfield  {journal} {\bibinfo
  {journal} {Phys. Rev. B}\ }\textbf {\bibinfo {volume} {87}},\ \bibinfo
  {pages} {125401} (\bibinfo {year} {2013}{\natexlab{b}})}\BibitemShut
  {NoStop}%
\bibitem [{\citenamefont {J.-J.~Zhu}\ and\ \citenamefont
  {Chang}(2011)}]{JJZhu}%
  \BibitemOpen
  \bibfield  {author} {\bibinfo {author} {\bibfnamefont {S.-C.~Z.}\
  \bibnamefont {J.-J.~Zhu}, \bibfnamefont {D.-X.~Yao}}\ and\ \bibinfo {author}
  {\bibfnamefont {K.}~\bibnamefont {Chang}},\ }\href {\doibase
  10.1103/PhysRevLett.106.097201} {\bibfield  {journal} {\bibinfo  {journal}
  {Phys. Rev. Lett.}\ }\textbf {\bibinfo {volume} {106}},\ \bibinfo {pages}
  {097201} (\bibinfo {year} {2011})}\BibitemShut {NoStop}%
\bibitem [{\citenamefont {Hosseini}\ and\ \citenamefont
  {Askari}(2015)}]{M.V.Hosseini}%
  \BibitemOpen
  \bibfield  {author} {\bibinfo {author} {\bibfnamefont {M.~V.}\ \bibnamefont
  {Hosseini}}\ and\ \bibinfo {author} {\bibfnamefont {M.}~\bibnamefont
  {Askari}},\ }\href {\doibase 10.1103/PhysRevB.92.224435} {\bibfield
  {journal} {\bibinfo  {journal} {Phys. Rev. B}\ }\textbf {\bibinfo {volume}
  {92}},\ \bibinfo {pages} {224435} (\bibinfo {year} {2015})}\BibitemShut
  {NoStop}%
\bibitem [{\citenamefont {Parhizgar}\ \emph
  {et~al.}(2013{\natexlab{c}})\citenamefont {Parhizgar}, \citenamefont
  {Asgari}, \citenamefont {Abedinpour},\ and\ \citenamefont
  {Zareyan}}]{fariborz-sp}%
  \BibitemOpen
  \bibfield  {author} {\bibinfo {author} {\bibfnamefont {F.}~\bibnamefont
  {Parhizgar}}, \bibinfo {author} {\bibfnamefont {R.}~\bibnamefont {Asgari}},
  \bibinfo {author} {\bibfnamefont {S.}~\bibnamefont {Abedinpour}}, \ and\
  \bibinfo {author} {\bibfnamefont {M.}~\bibnamefont {Zareyan}},\ }\href
  {\doibase 10.1103/PhysRevB.87.125402} {\bibfield  {journal} {\bibinfo
  {journal} {Phys. Rev. B}\ }\textbf {\bibinfo {volume} {12}},\ \bibinfo
  {pages} {125402} (\bibinfo {year} {2013}{\natexlab{c}})}\BibitemShut
  {NoStop}%
\bibitem [{\citenamefont {Sherafati}\ and\ \citenamefont
  {Satpathy}(2011)}]{sherafati-g}%
  \BibitemOpen
  \bibfield  {author} {\bibinfo {author} {\bibfnamefont {M.}~\bibnamefont
  {Sherafati}}\ and\ \bibinfo {author} {\bibfnamefont {S.}~\bibnamefont
  {Satpathy}},\ }\href {\doibase 10.1103/PhysRevB.83.165425} {\bibfield
  {journal} {\bibinfo  {journal} {Phys. Rev. B}\ }\textbf {\bibinfo {volume}
  {83}},\ \bibinfo {pages} {165425} (\bibinfo {year} {2011})}\BibitemShut
  {NoStop}%
\bibitem [{\citenamefont {Shiranzaei}\ \emph
  {et~al.}(2017{\natexlab{b}})\citenamefont {Shiranzaei}, \citenamefont
  {Parhizgar}, \citenamefont {Fransson},\ and\ \citenamefont
  {Cheraghchi}}]{Mahroo-single}%
  \BibitemOpen
  \bibfield  {author} {\bibinfo {author} {\bibfnamefont {M.}~\bibnamefont
  {Shiranzaei}}, \bibinfo {author} {\bibfnamefont {F.}~\bibnamefont
  {Parhizgar}}, \bibinfo {author} {\bibfnamefont {J.}~\bibnamefont {Fransson}},
  \ and\ \bibinfo {author} {\bibfnamefont {H.}~\bibnamefont {Cheraghchi}},\
  }\href {\doibase 10.1103/PhysRevB.95.235429} {\bibfield  {journal} {\bibinfo
  {journal} {Phys. Rev. B}\ }\textbf {\bibinfo {volume} {95}},\ \bibinfo
  {pages} {235429} (\bibinfo {year} {2017}{\natexlab{b}})}\BibitemShut
  {NoStop}%
\bibitem [{\citenamefont {M.~Zare}(2018)}]{Moslem18}%
  \BibitemOpen
  \bibfield  {author} {\bibinfo {author} {\bibfnamefont {R.~A.}\ \bibnamefont
  {M.~Zare}, \bibfnamefont {F.~Parhizgar}},\ }\href {\doibase
  10.1016/j.jmmm.2018.02.049} {\bibfield  {journal} {\bibinfo  {journal} {J.
  Magn. Magn. Mater.}\ }\textbf {\bibinfo {volume} {456}},\ \bibinfo {pages}
  {307} (\bibinfo {year} {2018})}\BibitemShut {NoStop}%
\bibitem [{\citenamefont {Zuti\'{c}}\ \emph {et~al.}(2004)\citenamefont
  {Zuti\'{c}}, \citenamefont {Fabian},\ and\ \citenamefont {Sarma}}]{fabian}%
  \BibitemOpen
  \bibfield  {author} {\bibinfo {author} {\bibfnamefont {I.}~\bibnamefont
  {Zuti\'{c}}}, \bibinfo {author} {\bibfnamefont {J.}~\bibnamefont {Fabian}}, \
  and\ \bibinfo {author} {\bibfnamefont {S.~D.}\ \bibnamefont {Sarma}},\ }\href
  {\doibase 10.1103/RevModPhys.76.323} {\bibfield  {journal} {\bibinfo
  {journal} {Rev. Mod. Phys.}\ }\textbf {\bibinfo {volume} {73}},\ \bibinfo
  {pages} {323} (\bibinfo {year} {2004})}\BibitemShut {NoStop}%
\bibitem [{\citenamefont {Babar}\ and\ \citenamefont {Kabir}(2016)}]{Babar}%
  \BibitemOpen
  \bibfield  {author} {\bibinfo {author} {\bibfnamefont {R.}~\bibnamefont
  {Babar}}\ and\ \bibinfo {author} {\bibfnamefont {M.}~\bibnamefont {Kabir}},\
  }\href {\doibase 10.1021/acs.jpcc.6b05069} {\bibfield  {journal} {\bibinfo
  {journal} {J. Phys. Chem. C}\ }\textbf {\bibinfo {volume} {120}},\ \bibinfo
  {pages} {27} (\bibinfo {year} {2016})}\BibitemShut {NoStop}%
\bibitem [{\citenamefont {W.~Han}(2014)}]{W.Han}%
  \BibitemOpen
  \bibfield  {author} {\bibinfo {author} {\bibfnamefont {M.~G.-J.~F.}\
  \bibnamefont {W.~Han}, \bibfnamefont {R.~K.~Kawakami}},\ }\href {\doibase
  10.1063/1.4895924} {\bibfield  {journal} {\bibinfo  {journal} {Nat. Nano.}\
  }\textbf {\bibinfo {volume} {9}},\ \bibinfo {pages} {794–807} (\bibinfo
  {year} {2014})}\BibitemShut {NoStop}%
\bibitem [{\citenamefont {Zare}\ and\ \citenamefont
  {Sadeghi}(2018)}]{MoslemBP}%
  \BibitemOpen
  \bibfield  {author} {\bibinfo {author} {\bibfnamefont {M.}~\bibnamefont
  {Zare}}\ and\ \bibinfo {author} {\bibfnamefont {E.}~\bibnamefont {Sadeghi}},\
  }\href {\doibase 10.1103/PhysRevB.98.205401} {\bibfield  {journal} {\bibinfo
  {journal} {Phys. Rev. B}\ }\textbf {\bibinfo {volume} {98}},\ \bibinfo
  {pages} {205401} (\bibinfo {year} {2018})}\BibitemShut {NoStop}%
\bibitem [{\citenamefont {Klinovaja}\ and\ \citenamefont
  {Loss}(2013)}]{Klinovaja13}%
  \BibitemOpen
  \bibfield  {author} {\bibinfo {author} {\bibfnamefont {J.}~\bibnamefont
  {Klinovaja}}\ and\ \bibinfo {author} {\bibfnamefont {D.}~\bibnamefont
  {Loss}},\ }\href {\doibase 10.1103/PhysRevB.87.045422} {\bibfield  {journal}
  {\bibinfo  {journal} {Phys. Rev. B}\ }\textbf {\bibinfo {volume} {87}},\
  \bibinfo {pages} {045422} (\bibinfo {year} {2013})}\BibitemShut {NoStop}%
\bibitem [{\citenamefont {H.~Duan}\ and\ \citenamefont
  {R.-Q.Wang}(2017)}]{Duan17}%
  \BibitemOpen
  \bibfield  {author} {\bibinfo {author} {\bibfnamefont {S.-H. Z.-Z. S. M.~Y.}\
  \bibnamefont {H.~Duan}, \bibfnamefont {S.~Li}}\ and\ \bibinfo {author}
  {\bibnamefont {R.-Q.Wang}},\ }\href {\doibase 10.1088/1367-2630/aa833a}
  {\bibfield  {journal} {\bibinfo  {journal} {New J. Phys.}\ }\textbf {\bibinfo
  {volume} {19}},\ \bibinfo {pages} {103010} (\bibinfo {year}
  {2017})}\BibitemShut {NoStop}%
\bibitem [{\citenamefont {Pereira}\ and\ \citenamefont
  {Neto}(2009)}]{Pereira09}%
  \BibitemOpen
  \bibfield  {author} {\bibinfo {author} {\bibfnamefont {V.~M.}\ \bibnamefont
  {Pereira}}\ and\ \bibinfo {author} {\bibfnamefont {A.~H.~C.}\ \bibnamefont
  {Neto}},\ }\href {\doibase 10.1103/PhysRevLett.103.046801} {\bibfield
  {journal} {\bibinfo  {journal} {Phys. Rev. Lett.}\ }\textbf {\bibinfo
  {volume} {103}},\ \bibinfo {pages} {046801} (\bibinfo {year}
  {2009})}\BibitemShut {NoStop}%
\bibitem [{\citenamefont {F.~Liu}\ and\ \citenamefont
  {Li}(2007)}]{PhysRevB.76.064120}%
  \BibitemOpen
  \bibfield  {author} {\bibinfo {author} {\bibfnamefont {P.~M.}\ \bibnamefont
  {F.~Liu}}\ and\ \bibinfo {author} {\bibfnamefont {J.}~\bibnamefont {Li}},\
  }\href@noop {} {\bibfield  {journal} {\bibinfo  {journal} {Phys. Rev. B}\
  }\textbf {\bibinfo {volume} {76}},\ \bibinfo {pages} {064120} (\bibinfo
  {year} {2007})}\BibitemShut {NoStop}%
\bibitem [{\citenamefont {V.~M.~Pereira}\ and\ \citenamefont
  {Peres}(2009)}]{pereira_tight-binding_2009}%
  \BibitemOpen
  \bibfield  {author} {\bibinfo {author} {\bibfnamefont {A.~H. C.~N.}\
  \bibnamefont {V.~M.~Pereira}}\ and\ \bibinfo {author} {\bibfnamefont
  {N.~M.~R.}\ \bibnamefont {Peres}},\ }\href {\doibase
  10.1103/PhysRevB.80.045401} {\bibfield  {journal} {\bibinfo  {journal} {Phys.
  Rev. B}\ }\textbf {\bibinfo {volume} {80}},\ \bibinfo {pages} {045401}
  (\bibinfo {year} {2009})}\BibitemShut {NoStop}%
\bibitem [{\citenamefont {S.~R.~Power}\ and\ \citenamefont
  {Ferreira}(2012)}]{ourpaper}%
  \BibitemOpen
  \bibfield  {author} {\bibinfo {author} {\bibfnamefont {J.~M.~D.}\
  \bibnamefont {S.~R.~Power}, \bibfnamefont {P.~D.~Gorman}}\ and\ \bibinfo
  {author} {\bibfnamefont {M.~S.}\ \bibnamefont {Ferreira}},\ }\href {\doibase
  10.1103/PhysRevB.86.195423} {\bibfield  {journal} {\bibinfo  {journal} {Phys.
  Rev. B}\ }\textbf {\bibinfo {volume} {86}},\ \bibinfo {pages} {195423}
  (\bibinfo {year} {2012})}\BibitemShut {NoStop}%
\bibitem [{\citenamefont {Peng}\ and\ \citenamefont
  {Hongbin}(2012)}]{Peng20123434}%
  \BibitemOpen
  \bibfield  {author} {\bibinfo {author} {\bibfnamefont {F.}~\bibnamefont
  {Peng}}\ and\ \bibinfo {author} {\bibfnamefont {W.}~\bibnamefont {Hongbin}},\
  }\href {\doibase 10.1016/j.physb.2012.04.053} {\bibfield  {journal} {\bibinfo
   {journal} {Phys. B (Amsterdam)}\ }\textbf {\bibinfo {volume} {407}},\
  \bibinfo {pages} {3434} (\bibinfo {year} {2012})}\BibitemShut {NoStop}%
\bibitem [{\citenamefont {F.~Guinea}\ and\ \citenamefont
  {Geim}(2010)}]{Guinea:gapsgraphene}%
  \BibitemOpen
  \bibfield  {author} {\bibinfo {author} {\bibfnamefont {M.~I.~K.}\
  \bibnamefont {F.~Guinea}}\ and\ \bibinfo {author} {\bibfnamefont {A.~K.}\
  \bibnamefont {Geim}},\ }\href {\doibase 10.1038/nphys1420} {\bibfield
  {journal} {\bibinfo  {journal} {Nat. Phys.}\ }\textbf {\bibinfo {volume}
  {6}},\ \bibinfo {pages} {30} (\bibinfo {year} {2010})}\BibitemShut {NoStop}%
\bibitem [{\citenamefont {A.~Sharma}\ and\ \citenamefont
  {Neto}(2013)}]{sharma_effect_2013}%
  \BibitemOpen
  \bibfield  {author} {\bibinfo {author} {\bibfnamefont {V.~N.~K.}\
  \bibnamefont {A.~Sharma}}\ and\ \bibinfo {author} {\bibfnamefont {A.~H.~C.}\
  \bibnamefont {Neto}},\ }\href {\doibase 10.1103/PhysRevB.87.155431}
  {\bibfield  {journal} {\bibinfo  {journal} {Phys. Rev. B}\ }\textbf {\bibinfo
  {volume} {87}},\ \bibinfo {pages} {155431} (\bibinfo {year}
  {2013})}\BibitemShut {NoStop}%
\bibitem [{\citenamefont {Y.~Zhao}\ and\ \citenamefont {Wang}(2018)}]{Yu.Zhao}%
  \BibitemOpen
  \bibfield  {author} {\bibinfo {author} {\bibfnamefont {J.~L. C.~Z.}\
  \bibnamefont {Y.~Zhao}, \bibfnamefont {X.~Li}}\ and\ \bibinfo {author}
  {\bibfnamefont {Q.}~\bibnamefont {Wang}},\ }\href {\doibase
  10.1021/acs.jpclett.8b00616} {\bibfield  {journal} {\bibinfo  {journal} {J.
  Phys. Chem. Lett.}\ }\textbf {\bibinfo {volume} {9}},\ \bibinfo {pages}
  {1815} (\bibinfo {year} {2018})}\BibitemShut {NoStop}%
\bibitem [{\citenamefont {P.~Li}\ and\ \citenamefont {Yang}(2018)}]{P.Li}%
  \BibitemOpen
  \bibfield  {author} {\bibinfo {author} {\bibfnamefont {Z.~L.}\ \bibnamefont
  {P.~Li}}\ and\ \bibinfo {author} {\bibfnamefont {J.}~\bibnamefont {Yang}},\
  }\href {\doibase 10.1021/acs.jpclett.8b02035} {\bibfield  {journal} {\bibinfo
   {journal} {J. Phys. Chem. Lett.}\ }\textbf {\bibinfo {volume} {9}},\
  \bibinfo {pages} {4852–4856} (\bibinfo {year} {2018})}\BibitemShut
  {NoStop}%
\bibitem [{\citenamefont {L.~Z.~Zhang}(2014)}]{L.Z.Zhang}%
  \BibitemOpen
  \bibfield  {author} {\bibinfo {author} {\bibfnamefont {S.~X. D. H. J. G.
  F.~L.}\ \bibnamefont {L.~Z.~Zhang}, \bibfnamefont {Z.~F.~Wang}},\ }\href
  {\doibase 10.1103/PhysRevB.90.161402} {\bibfield  {journal} {\bibinfo
  {journal} {Phys. Rev. B}\ }\textbf {\bibinfo {volume} {90}},\ \bibinfo
  {pages} {161402(R)} (\bibinfo {year} {2014})}\BibitemShut {NoStop}%
\bibitem [{\citenamefont {H.~Zhang}(2016)}]{H.Zhang}%
  \BibitemOpen
  \bibfield  {author} {\bibinfo {author} {\bibfnamefont {J.~H. A. D. Z.~C.}\
  \bibnamefont {H.~Zhang}, \bibfnamefont {Y.~Li}},\ }\href@noop {} {\bibfield
  {journal} {\bibinfo  {journal} {Nano Lett.}\ }\textbf {\bibinfo {volume}
  {16}},\ \bibinfo {pages} {6124} (\bibinfo {year} {2016})}\BibitemShut
  {NoStop}%
\bibitem [{\citenamefont {J.~Qiao}\ and\ \citenamefont {Ji}(2014)}]{}%
  \BibitemOpen
  \bibfield  {author} {\bibinfo {author} {\bibfnamefont {Z.-X. H. F.~Y.}\
  \bibnamefont {J.~Qiao}, \bibfnamefont {X.~Kong}}\ and\ \bibinfo {author}
  {\bibfnamefont {W.}~\bibnamefont {Ji}},\ }\href {\doibase 10.1038/ncomms5475
  (2014)} {\bibfield  {journal} {\bibinfo  {journal} {Nat. Commun.}\ }\textbf
  {\bibinfo {volume} {5}},\ \bibinfo {pages} {4475} (\bibinfo {year}
  {2014})}\BibitemShut {NoStop}%
\bibitem [{\citenamefont {Harrison}(1999)}]{HarrisonWA1999}%
  \BibitemOpen
  \bibfield  {author} {\bibinfo {author} {\bibfnamefont {W.~A.}\ \bibnamefont
  {Harrison}},\ }\href@noop {} {\  (\bibinfo {year} {1999})}\BibitemShut
  {NoStop}%
\bibitem [{\citenamefont {H.~Tang}\ and\ \citenamefont {Su}(2009)}]{TangH}%
  \BibitemOpen
  \bibfield  {author} {\bibinfo {author} {\bibfnamefont {B.~S.~W.}\
  \bibnamefont {H.~Tang}, \bibfnamefont {J.~W.Jiang}}\ and\ \bibinfo {author}
  {\bibfnamefont {Z.~B.}\ \bibnamefont {Su}},\ }\href {\doibase
  10.1016/j.ssc.2008.10.012} {\bibfield  {journal} {\bibinfo  {journal} {Solid
  State Commun.}\ }\textbf {\bibinfo {volume} {149}},\ \bibinfo {pages} {82}
  (\bibinfo {year} {2009})}\BibitemShut {NoStop}%
\bibitem [{\citenamefont {Jiang}\ and\ \citenamefont {~}(2015)}]{J.W.Jiang}%
  \BibitemOpen
  \bibfield  {author} {\bibinfo {author} {\bibfnamefont {J.~W.}\ \bibnamefont
  {Jiang}}\ and\ \bibinfo {author} {\bibfnamefont {H.~S.~P.}\ \bibnamefont
  {~}},\ }\href {\doibase 10.1103/PhysRevB.91.235118} {\bibfield  {journal}
  {\bibinfo  {journal} {Phys. Rev. B}\ }\textbf {\bibinfo {volume} {91}},\
  \bibinfo {pages} {235118} (\bibinfo {year} {2015})}\BibitemShut {NoStop}%
\bibitem [{\citenamefont {Lu}\ and\ \citenamefont {Guo}(2010)}]{Y.Lu2010}%
  \BibitemOpen
  \bibfield  {author} {\bibinfo {author} {\bibfnamefont {Y.}~\bibnamefont
  {Lu}}\ and\ \bibinfo {author} {\bibfnamefont {J.}~\bibnamefont {Guo}},\
  }\href {\doibase 10.1007/s12274-010-1022-4} {\bibfield  {journal} {\bibinfo
  {journal} {Nano Res}\ }\textbf {\bibinfo {volume} {3}},\ \bibinfo {pages}
  {189–199} (\bibinfo {year} {2010})}\BibitemShut {NoStop}%
\bibitem [{\citenamefont {Y.~Zhang}\ and\ \citenamefont
  {Yang}(2012)}]{Y.Zhang12}%
  \BibitemOpen
  \bibfield  {author} {\bibinfo {author} {\bibfnamefont {Q.~L.}\ \bibnamefont
  {Y.~Zhang}, \bibfnamefont {X.~Wu}}\ and\ \bibinfo {author} {\bibfnamefont
  {J.}~\bibnamefont {Yang}},\ }\href {\doibase 10.1021/jp301691z} {\bibfield
  {journal} {\bibinfo  {journal} {J. Phys. Chem. C 116, 16,}\ }\textbf
  {\bibinfo {volume} {116 (16)}},\ \bibinfo {pages} {9356} (\bibinfo {year}
  {2012})}\BibitemShut {NoStop}%
\bibitem [{\citenamefont {A.~Carvalho}(2014)}]{A.CarvalhoEPL}%
  \BibitemOpen
  \bibfield  {author} {\bibinfo {author} {\bibfnamefont {A.~N.}\ \bibnamefont
  {A.~Carvalho}, \bibfnamefont {A.~Rodin}},\ }\href {\doibase
  10.1209/0295-5075/108/47005} {\bibfield  {journal} {\bibinfo  {journal}
  {Europhys. Lett. (EPL)}\ }\textbf {\bibinfo {volume} {108 (4)}},\ \bibinfo
  {pages} {47005} (\bibinfo {year} {2014})}\BibitemShut {NoStop}%
\bibitem [{\citenamefont {H.~Guo}\ and\ \citenamefont {Zeng}(2014)}]{H.Guo14}%
  \BibitemOpen
  \bibfield  {author} {\bibinfo {author} {\bibfnamefont {J.~D. X.~W.}\
  \bibnamefont {H.~Guo}, \bibfnamefont {N.~Lu}}\ and\ \bibinfo {author}
  {\bibfnamefont {X.~C.}\ \bibnamefont {Zeng}},\ }\href {\doibase
  10.1021/jp505257g} {\bibfield  {journal} {\bibinfo  {journal} {J. Phys. Chem.
  C}\ }\textbf {\bibinfo {volume} {118 (25)}},\ \bibinfo {pages} {14051}
  (\bibinfo {year} {2014})}\BibitemShut {NoStop}%
\bibitem [{\citenamefont {Soleimanikahnoj}\ and\ \citenamefont
  {Knezevic}(2017)}]{Soleimanikahnoj}%
  \BibitemOpen
  \bibfield  {author} {\bibinfo {author} {\bibfnamefont {S.}~\bibnamefont
  {Soleimanikahnoj}}\ and\ \bibinfo {author} {\bibfnamefont {I.}~\bibnamefont
  {Knezevic}},\ }\href {\doibase 10.1103/PhysRevApplied.8.064021} {\bibfield
  {journal} {\bibinfo  {journal} {Phys. Rev. Appl.}\ }\textbf {\bibinfo
  {volume} {8}},\ \bibinfo {pages} {064021} (\bibinfo {year}
  {2017})}\BibitemShut {NoStop}%
\bibitem [{\citenamefont {Imamura}\ \emph {et~al.}(2004)\citenamefont
  {Imamura}, \citenamefont {Bruno},\ and\ \citenamefont {Utsumi}}]{Imamura}%
  \BibitemOpen
  \bibfield  {author} {\bibinfo {author} {\bibfnamefont {H.}~\bibnamefont
  {Imamura}}, \bibinfo {author} {\bibfnamefont {P.}~\bibnamefont {Bruno}}, \
  and\ \bibinfo {author} {\bibfnamefont {Y.}~\bibnamefont {Utsumi}},\ }\href
  {\doibase 10.1103/PhysRevB.69.121303} {\bibfield  {journal} {\bibinfo
  {journal} {Phys. Rev. B}\ }\textbf {\bibinfo {volume} {69}},\ \bibinfo
  {pages} {121303(R)} (\bibinfo {year} {2004})}\BibitemShut {NoStop}%
\end{thebibliography}%
\end{document}